\documentclass[preprint,prd,aps,nofootinbib,onecolumn,11pt]{revtex4}
\usepackage{amsmath}
\usepackage{epsfig}
\usepackage{amssymb}
\usepackage{subfigure}
\usepackage{color}
\usepackage{slashed}
\usepackage{verbatim}
\begin{document}

\preprint{SI-HEP-2017-25}
\preprint{QFET-2017-22}

\title{Charged lepton flavor violating Higgs decays at future $e^+e^-$ colliders}
\author{Qin Qin$^{1}$}\email{qin@physik.uni-siegen.de}
\author{Qiang Li$^{2,3}$}\email{qliphy0@pku.edu.cn}
\author{Cai-Dian L\"u$^{3,4}$}\email{lucd@ihep.ac.cn}
\author{Fu-Sheng Yu$^{5,6}$}\email{yufsh@lzu.edu.cn}
\author{Si-Hong Zhou$^{4,7}$}\email{shzhou@ihep.ac.cn}
\affiliation{$^1$Theoretische Physik 1, Naturwissenschaftlich-Technische Fakult\"at, Universit\"at Siegen, Walter-Flex-Strasse 3, D-57068 Siegen, Germany}
\affiliation{$^2$Department of Physics and State Key Laboratory of Nuclear Physics and Technology, Peking University, Beijing, 100871, China}
\affiliation{$^3$CAS Center for Excellence in Particle Physics, Beijing 100049, China}
\affiliation{$^4$Institute of High Energy Physics, P.O. Box 918(4), Chinese Academy of Sciences, Beijing 100049, People's Republic of China}
\affiliation{$^5$School of Nuclear Science and Technology, Lanzhou University, Lanzhou 730000, People's Republic of China}
\affiliation{$^6$Research Center for Hadron and CSR Physics, Lanzhou University and Institute of Modern Physics of CAS, Lanzhou 730000, People's Republic of China}
\affiliation{$^7$School of Physical Science and Technology, Inner Mongolia University, Hohhot 010021, China}

\begin{abstract}
After the discovery of the Higgs boson, several future experiments have been proposed to study the Higgs 
boson properties, including two circular lepton colliders, the CEPC and the FCC-ee, and one linear lepton collider, the ILC.
We evaluate the precision reach of these colliders in measuring the branching ratios of the charged lepton 
flavor violating Higgs decays $H\to e^\pm\mu^\mp$, $e^\pm\tau^\mp$ and $\mu^\pm\tau^\mp$. The expected 
upper bounds on the branching ratios given by the circular (linear) colliders are 
found to be $\mathcal{B}(H\to e^\pm\mu^\mp) < 1.2\ (2.1) \times 10^{-5}$, $\mathcal{B}(H\to e^\pm\tau^\mp) < 1.6\ (2.4) \times 10^{-4}$ 
and $\mathcal{B}(H\to \mu^\pm\tau^\mp) < 1.4\ (2.3) \times 10^{-4}$ at 95\% CL, which are improved 
by one to two orders compared to the current experimental bounds. We also 
discuss the constraints that these upper bounds set on certain theory parameters, including the 
charged lepton flavor violating Higgs couplings, the corresponding parameters in the type-III 2HDM, 
and the new physics cut-off scales in the SMEFT, in RS models and in models with heavy neutrinos.
\end{abstract}

%%%%%%%%%%%%%%%%%%%%%%%%%%%%%%%%%%%%%%%%%%%%%%%%%%%%%%%%%%%%%%%%%%%%%%%%%%%%%%%%%%%%%%%%%%%%%%%%%%%%%%%%%%%%%%%%%%%%%
\maketitle

\section{Introduction}
  
The discovery of the Higgs boson \cite{Aad:2012tfa,Chatrchyan:2012xdj} not only completes the 
Standard Model (SM) of particle physics, but also opens new windows for searches for new physics 
through the Higgs portal. Although current experimental results \cite{Chatrchyan:2013lba,
Aad:2013wqa,Aad:2013xqa,Khachatryan:2014jba,Khachatryan:2014kca,Aad:2015zhl} indicate a
preference for a SM-like Higgs boson, more precise measurements are required in order to determine 
its true nature and whether or not it has new physics properties. 
Three electron-positron colliders, the Circular Electron-Positron Collider (CEPC) \cite{CEPCStudyGroup:2018rmc},  
the FCC-ee, formerly known as TLEP \cite{Gomez-Ceballos:2013zzn}, and the International Linear 
Collider (ILC) \cite{Baer:2013cma}, have been proposed by different high-energy communities, 
aiming to precisely study the Higgs boson properties. They are designed to operate at 240 - 250 GeV 
with a large sample of Higgs bosons collected, mainly by the $e^+e^-\to ZH$ process. 
The large amount of Higgs bosons produced in a clean environment will allow measurements of the 
cross section of the Higgs production \cite{Mo:2015mza} as well as its mass \cite{Chen:2016zpw,Xu:2015goa,Gong:2016jys}, 
decay width \cite{Chen:2016prx} and branching ratios \cite{Liu:2016zki,Chen:2016zpw,Chen:2017ipx,Cui:2017dqp} with 
precision far beyond that of the Large Hadron Collider (LHC). Such machines will also provide opportunities  to search for 
new particles such as new multi-quark states \cite{Ali:2018ifm}, dark photons \cite{He:2017ord}, dark matter particles 
\cite{Cao:2016qgc,Liu:2017lpo,Cai:2017wdu,Xiang:2017yfs,Wang:2017sxx}, 
heavy neutrinos \cite{Antusch:2015rma,Antusch:2017pkq,Liao:2017jiz,Banerjee:2015gca} and supersymmetric particles 
\cite{Potter:2017ajd}, and also to probe new physics scales via Higgs and electroweak observables 
\cite{Fan:2014vta,Fan:2014axa,Fedderke:2015txa,Huang:2015izx,Ge:2016zro,Durieux:2017rsg,
Gori:2015nqa,Craig:2015wwr,Li:2016zzh}. In this work, we focus on the charged lepton flavor violating 
(CLFV) Higgs decays $H\to e^\pm\mu^\mp$, $e^\pm\tau^\mp$ and $\mu^\pm\tau^\mp$.

The CLFV Higgs decays are interesting, because their observation may provide insight into some fundamental 
questions in nature, {\it e.g.}, whether there is a secondary mechanism for the electroweak symmetry breaking \cite{DiazCruz:1999xe}, 
why the neutrino masses are tiny \cite{ArkaniHamed:2000bq}, and whether there is an extra 
dimension responsible for the gauge hierarchy generation \cite{Perez:2008ee}. They have thus 
attracted a lot of attention from both theorists and experimentalists \cite{Bjorken:1977vt,McWilliams:1980kj,Han:2000jz,
Arganda:2004bz,Goudelis:2011un,Arhrib:2012mg,Azatov:2009na,Blankenburg:2012ex,Arganda:2014dta,
Arganda:2015naa,Arganda:2015uca,Huang:2015vpt,Crivellin:2015mga,Beneke:2015lba,Baek:2015fma,
Feldmann:2016hvo,Han:2016bvl,Thuc:2016qva,Khachatryan:2015kon,Aad:2016blu,CMS:2016qvi}. 
The CMS collaboration reported the first hint of charged lepton flavor violation in the $H\to\mu^\pm\tau^\mp$ 
channel with a significance of 2.4 standard deviations \cite{Aad:2016blu,CMS:2016qvi}. Although this signal 
disappeared later \cite{Aad:2016blu,CMS:2016qvi}, the CLFV Higgs decays are 
still worthy to be studied with higher precision. On one hand, the so-called flavor anomalies, which indicate lepton 
flavor non-universality, reported by the B factories and the LHCb collaboration \cite{Lees:2012xj,Aaij:2014ora,
Huschle:2015rga,Aaij:2017vbb}, in some sense also imply lepton flavor violation (when the 
lepton mass matrices are diagonalised to obtain the physical states, unequal diagonal couplings with different 
leptons will lead to off-diagonal couplings). On the other hand, the $B$ meson decay channels in which the 
flavor anomalies are observed are always polluted by complicated strong dynamics, while the much cleaner 
CLFV Higgs decay channels will provide a better chance to study the mechanism generating the lepton 
flavor violation or non-universality once they are discovered. The potential to search for such decay channels 
at the High Luminosity LHC (HL-LHC) has been estimated in \cite{Banerjee:2016foh}. 
By this paper, we study the sensitivity of the three lepton colliders in measuring the CLFV Higgs decays based 
on the detector simulation of the signal events of the decay channels and the corresponding background. 
There are already some studies on the ILC measurement of $H\to\mu^\pm\tau^\mp$
\cite{Banerjee:2016foh,Kanemura:2004cn,Chakraborty:2016gff,Chakraborty:2017tyb}, and the difference between 
our paper and theirs will be discussed.

The paper is organized as follows: in Section \ref{sec:simulation}, we perform the detector simulation of the 
signal and background events for the three CLFV Higgs decay channels at the CEPC and obtain the upper 
bounds on the three decay rates, based on which we estimate the corresponding upper bounds expected to 
be given by the FCC-ee and the ILC; we derive in Section \ref{sec:theory} the constraints on theory 
parameters including the CLFV Higgs couplings, the relevant parameters in the type-III two-Higgs-doublet-model 
(2HDM) and the new physics cut-off scales in the SM effective field 
theory (SMEFT), in Randall-Sundrum (RS) models and in models with heavy neutrinos; we summarize by 
Section \ref{sec:summary}.

\section{Simulation and Analysis}\label{sec:simulation}

In this section, we evaluate the possible reach of the three colliders with $\sqrt{s}$ = 240 - 250 GeV, the CEPC,
the FCC-ee and the ILC, in measuring the branching ratios of the three CLFV Higgs decays. Investigating the 
dominant Higgs production process $e^+e^-\to ZH$, with considered only the $Z$ boson decaying hadronically 
into a quark pair, we expect that the signal events each contain two charged leptons of different flavors and two 
jets. Therefore, the four-fermion processes will form the major SM background. The signal processes
$e^+e^-\to Z(\to q\bar{q})H(\to e^\pm\mu^\mp,e^\pm\tau^\mp,\mu^\pm\tau^\mp)$ are simulated via MadGraph 
v2.5.2 \cite{Alwall:2014hca}, with the corresponding three CLFV Higgs vertices implemented. The background 
events are generated via WHIZARD 2.5.0 \cite{Moretti:2001zz,Kilian:2007gr}. PYTHIA 6.4 \cite{Sjostrand:2006za} 
is then used to manage hadronization and parton showers for both the signal and background events. Finally, 
Delphes 3.4.1 \cite{Cacciari:2005hq,Cacciari:2011ma} is adopted for detector simulation. Note that we only 
carry out the above simulation procedure for the CEPC with $\sqrt{s}$ = 240 GeV and an integrated luminosity of 5 
ab$^{-1}$. We suppose that the FCC-ee with also $\sqrt{s}$ = 240 GeV has a same integrated luminosity and 
similar detector performance, and hence the CEPC results will apply to the FCC-ee. Different from the CEPC 
and the FCC-ee, the ILC is planned to run at $\sqrt{s}$ = 250 GeV with four polarization options: P1 (-0.8, 0.3),  
P2 (0.8, -0.3),  P3 (-0.8, -0.3) and  P4 (0.8, 0.3). The numbers give the degree of polarization of the beams. For 
example, the first option (P1) means that the $e^-$ beam is 80\% left-handed polarized and the $e^+$ beam is 
30\% right-handed polarized. The integrated luminosities for the four polarization options are (in fb$^{-1}$) 1350, 
450, 100 and 100, respectively. We find that the beam polarization will considerably change neither the angular 
distribution of the signal final states nor the statistical uncertainties owing to the background, which makes it 
possible to estimate the ILC sensitivity to the CLFV Higgs decays based on the CEPC simulation, as discussed 
in detail for each channel below in this section.

\begin{table}[!htbh]
  \centering
  \begin{tabular}[t]{|l|c|c|}\hline
 \parbox[0pt][2em][c]{0cm}{} \textbf{Category}\footnote{The four fermion processes are classified into different categories as follows: SZ$\_$ll includes $e^+e^-\tau^+\tau^-$;
  SZ$\_$ql includes $e^+e^-u\bar{u}$, $e^+e^-d\bar{d}$, $e^+e^-c\bar{c}$, $e^+e^-s\bar{s}$ and $e^+e^-b\bar{b}$; 
  ZZ$\_$ll includes $\mu^+\mu^-\tau^+\tau^-$ and $\tau^+\tau^-\tau^+\tau^-$;
  ZZ$\_$ql includes $\mu^+\mu^-u\bar{u}$, $\mu^+\mu^-d\bar{d}$, $\mu^+\mu^-c\bar{c}$, $\mu^+\mu^-s\bar{s}$, $\mu^+\mu^-b\bar{b}$, $\tau^+\tau^-u\bar{u}$, $\tau^+\tau^-d\bar{d}$, $\tau^+\tau^-c\bar{c}$, $\tau^+\tau^-s\bar{s}$ and $\tau^+\tau^-b\bar{b}$;
  ZZ$\_$qq includes $cc\bar{c}\bar{c}$, $dd\bar{d}\bar{d}$, $d\bar{d}b\bar{b}$, $bb\bar{b}\bar{b}$, $u\bar{u}s\bar{s}$, $u\bar{u}b\bar{b}$ and $s\bar{s}b\bar{b}$;
  SW$\_$ql includes $e^+\nu_e\bar{c}s$;
  WW$\_$ql includes $\tau^+\nu_\tau\bar{u}d$, $\mu^-\bar{\nu}_\mu c\bar{s}$ and $\mu^-\bar{\nu}_\mu u\bar{d}$.}  & \textbf{Cross section} [fb] & \textbf{Event} [$\times$10000]   \\ \hline
 \parbox[0pt][2em][c]{0cm}{} SZ$\_$ll & 342 & 171   \\ \hline
 \parbox[0pt][2em][c]{0cm}{} SZ$\_$ql & 452 & 226  \\ \hline
 \parbox[0pt][2em][c]{0cm}{} ZZ$\_$ll & 18.8 & 9.4    \\ \hline
 \parbox[0pt][2em][c]{0cm}{} ZZ$\_$ql & 233 & 117   \\ \hline
 \parbox[0pt][2em][c]{0cm}{} ZZ$\_$qq & 830 & 415    \\ \hline
 \parbox[0pt][2em][c]{0cm}{} SW$\_$ql & 667 & 333  \\ \hline
 \parbox[0pt][2em][c]{0cm}{} WW$\_$ql & 1792 & 896   \\ \hline
  \end{tabular}
  \caption{The cross sections and event numbers of the four fermion processes belonging to different categories at the 240 GeV CEPC with an integrated luminosity of 5 ab$^{-1}$. See text for details.}\label{tb:events}
\end{table}

For each CLFV Higgs decay channel, the production of 10000 signal events at the CEPC are simulated. 
We simulate the four fermion background events by WHIZARD with an integrated luminosity of 5 ab$^{-1}$, 
giving the cross sections and the numbers of events for different categories in Table \ref{tb:events}. 
The four fermion processes are classified into different 
categories according to their final states \cite{Bardin:1994sc} as follows. We divide the processes 
into two groups. The first group contains the processes without (anti-) electron or (anti-) electron neutrino in 
their final states, and the processes in the second group have at least one (anti-) electron or (anti-) electron 
neutrino in their final states. We start with a process whose final state is constituted by two pairs of mutually charge 
conjugate fermions like $u\bar{u}\mu^+\mu^-$ or $u\bar{u}e^+e^-$ that can not arise from decays of two $W$ bosons. If 
this process belongs to the first group, we will classify it into the "ZZ" category; if it belongs to the second group, 
it will be classified into "Single Z" (SZ). A process belonging to the first or the second group with two pairs of 
mutually charge conjugate fermions in the final state will be classified into the "ZZ or WW" (ZZWW) category or 
the "Single Z or Single W" (SZSW) category, respectively, if its final state can arise from decays of two $W$ bosons. The 
remaining processes in the first and second group are classified into "WW" and  "Single W" (SW), respectively. More details can 
be found in \cite{Mo:2015mza}. We further divide each category of processes into different sets, according to 
whether the final states contain only quarks ("qq"), only leptons ("ll") or both of them ("ql"). For example, 
$u\bar{u}\mu^+\mu^-$ is classified into "ZZ$\_$ql", while $\nu_\mu\bar{\nu}_\mu\mu^+\mu^-$ is classified into 
"ZZWW$\_$ll". Note we only consider the four fermion processes listed in the footnote of TABLE \ref{tb:events}. 
The neglected processes do not contribute to the background after the chosen cuts are made.

In the following, for each CLFV Higgs decay channel, we set appropriate cuts on both the generated signal and 
background events. Owing to the fact that the signal events of different channels need to be reconstructed differently 
and thus suffer from different backgrounds, different event selection cuts are determined through the significance 
optimization for different channels. Based on the signal detection efficiencies and the background event numbers, 
the CEPC bounds on the decay branching ratios are evaluated first, which also apply to the FCC-ee. After that, 
we estimate the corresponding ILC bounds by scaling the luminosity times the signal and background cross sections, 
\textit{i.e.}, scaling the event numbers, for each polarization option.

\subsection{$H\to e^\pm\mu^\mp$}

To reconstruct the signal events of $e^+e^-\to HZ \to e^\pm\mu^\mp Z$, we select the events with
final states containing one electron, one muon, and two jets to reconstruct the $Z$ boson, with 
the di-lepton (electron and muon) invariant mass $m_{e\mu}$ close to the Higgs boson mass and the di-jet
invariant mass $m_{jj}$ close to the $Z$ boson mass. This event selection condition requires the 
cuts 70 GeV $< m_{jj} <$ 100 GeV and 117 GeV $< m_{e\mu} <$ 127 GeV, which are displayed in 
TABLE \ref{tb:Higgsemu}. We use the di-jet to reconstruct the $Z$ boson because the large hadronic 
decay rate of the $Z$ boson $\mathcal{B}(Z\to q\bar{q})\approx$ 70\% \cite{Olive:2016xmw} ensures 
a high reconstruction efficiency. Besides, either if we choose an electron pair or a muon pair to 
reconstruct the $Z$ boson, leptons decaying from the Higgs boson and from the $Z$ boson may get mixed up.
Of course, one can always combine the analyses based on all the possible methods of reconstructing the 
$Z$ boson to improve the statistics, but we will not do this here as we only aim at an estimation of the 
order of the upper bounds on the CLFV decay rates in this work.

\begin{table}[!htbh]
  \centering
  \begin{tabular}[t]{|l|c|c|c|c|c|c|c|c|}\hline
 \parbox[0pt][2em][c]{0cm}{} \textbf{Cut} & SZ$\_$ll & SZ$\_$ql  & ZZ$\_$ll & ZZ$\_$ql  & ZZ$\_$qq & SW$\_$ql & WW$\_$ql & \textbf{Signal}  \\ \hline
 \parbox[0pt][2em][c]{0cm}{} $N_{e,\mu}$=1, $N_j$=2 & 5684 & 1248 & 1464 & 16504 & 1945 & 1063 & 1627 & 5617 \\ \hline
 \parbox[0pt][2em][c]{0cm}{} 70 GeV $< m_{jj} <$ 100 GeV & 1099 & 267 & 408 & 12277 & 321 & 221 & 461 & 4216  \\ \hline
 \parbox[0pt][2em][c]{0cm}{} 117 GeV $< m_{e\mu} <$ 127 GeV & 1 & 0  & 0 & 0 & 0 & 0 & 0 & 4115  \\ \hline
  \end{tabular}
  \caption{The numbers of background events in different categories and signal events surviving the cuts in the analysis of $H\to e^\pm\mu^\mp$ at the CEPC. $N_{e(\mu,j)}$ represents the number of electrons (muons, jets) in the final state of an event. See text for details.}\label{tb:Higgsemu}
\end{table}

In TABLE \ref{tb:Higgsemu} we list how many background events in different categories and how many 
signal events are left after the cuts based on the CEPC simulation. We find that 4115 out of the 10000 generated signal events 
survive the event selection, and thus the signal detection efficiency is $\epsilon$ = 41.15\%. As for 
the background, only one event is expected after the event selection. Given also one observed 
event as expected, the upper limit on the number of signal events (a Poisson variable) is $N_{95}$ 
= 3.74 at 95\% confidence level (CL). Then, employing 
\begin{equation}\label{eq:upperlimit}
\mathcal{B} < {N_{95}\over \epsilon N_{H}\mathcal{B}(Z\to q\bar{q})},
\end{equation}
with $\mathcal{B}(Z\to q\bar{q})\approx$ 70\% and $N_{H}$ = 1.05 $\times 10^6$ the number of the 
Higgs bosons to be produced by the CPEC, we evaluate the upper bound on the $H\to e^\pm\mu^\mp$ 
branching ratio, 
\begin{equation}
\mathcal{B}(H\to e^\pm\mu^\mp) < 1.2\times10^{-5}\ \text{at 95\% CL.}
\end{equation}
This bound is also accepted for the FCC-ee.

As for the ILC, the ratios of the $ZH$ production cross sections with the four polarization options 
to that of the CEPC are 
\begin{equation}
\sigma(ZH):\ \ \text{P1/CEPC = 1.48, P2/CEPC = 1.0, P3/CEPC = 0.87, P4/CEPC = 0.65.}
\end{equation}
Recalling the integrated luminosities for the four options, we obtain that about $5.21\times10^{5}$ 
Higgs bosons will be produced at the ILC. Through a tree-amplitude analysis of the 
$e^+e^-\to Z(\to jj)H(\to\ell^\pm\ell'^{\mp})$ process, we find that the beam polarization will not 
change the angular distribution of the final state, and thus the signal detection efficiency 
$\epsilon$ = 41.15\% is also valid under the same event selection condition. The only one 
background event listed in Table \ref{tb:Higgsemu} arises from the process $e^+e^-\tau^+\tau^-$. 
We calculate the its production cross section with each polarization option as what we did for the 
signal, and find that the background event number after 
2 ab$^{-1}$ data collected is expected to be 0.38. Then, for the ILC $N_{95}$ = 3.3 and 
the upper bound on $\mathcal{B}(H\to e^\pm\mu^\mp)$ is about 2.1$\times10^{-5}$ at 95\% CL.
The upper bound does get improved by introduction of beam polarization, mainly owing to the 
large $ZH$ production cross section of P1, but the improvement is not big enough to fill the gap 
of the integrated luminosities between the ILC and the CEPC (FCC-ee).
One might be unsatisfied with that we estimate the background by scaling the total event numbers, 
since the beam polarization might change the $e^+e^-\tau^+\tau^-$ angular distribution. We then 
test the polarization impact by assuming an extreme situation where the role of the polarization is 
so important that the background event number at the ILC is 0. Even in such a 
case, the upper bound is then only reduced to 1.9$\times10^{-5}$, by less than 10\% of magnitude. 
The key point is that the CEPC background is already very small, so further reduction of the 
background will not optimize the upper bound significantly. Therefore, we conclude that, regarded 
as estimates, our results for the ILC upper bounds (also for the other two channels) are acceptable.
 
\begin{figure}[ht!]
\centering
\includegraphics[width=70mm]{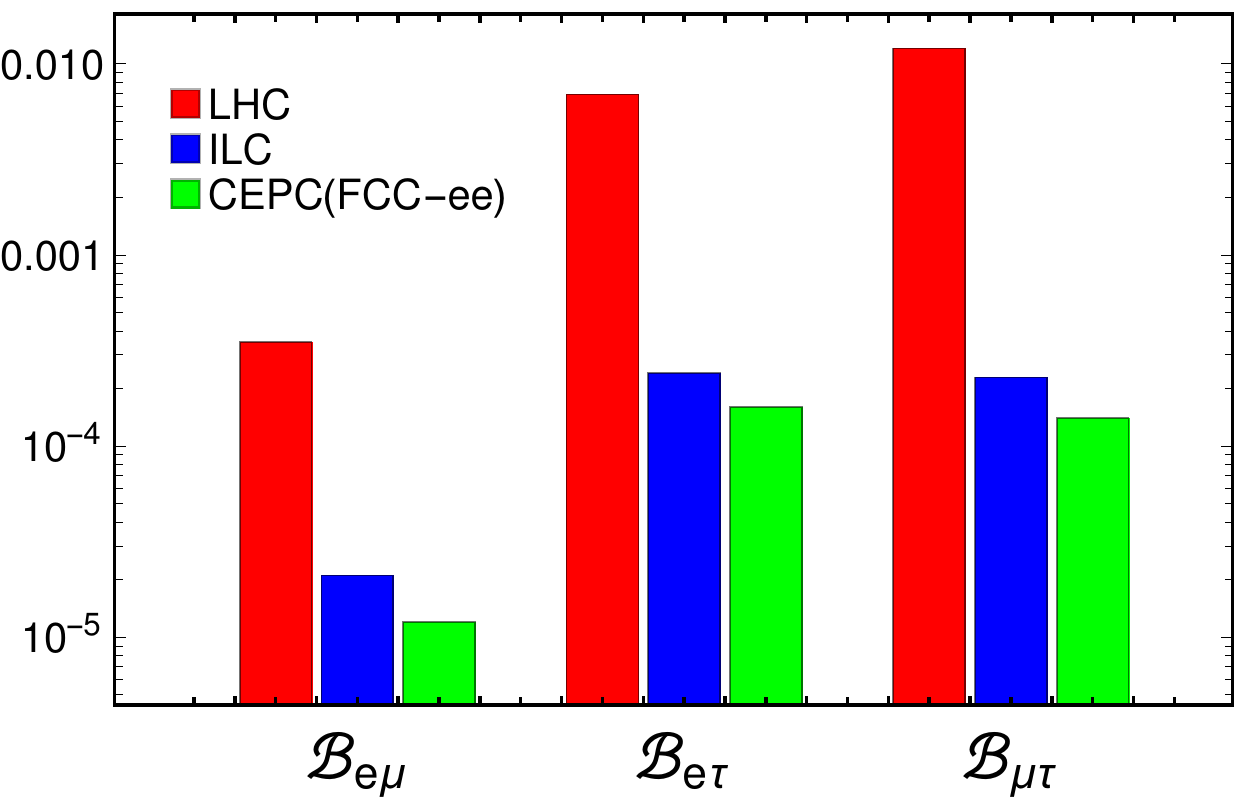}
  \hspace{0.2in}
\includegraphics[width=73mm]{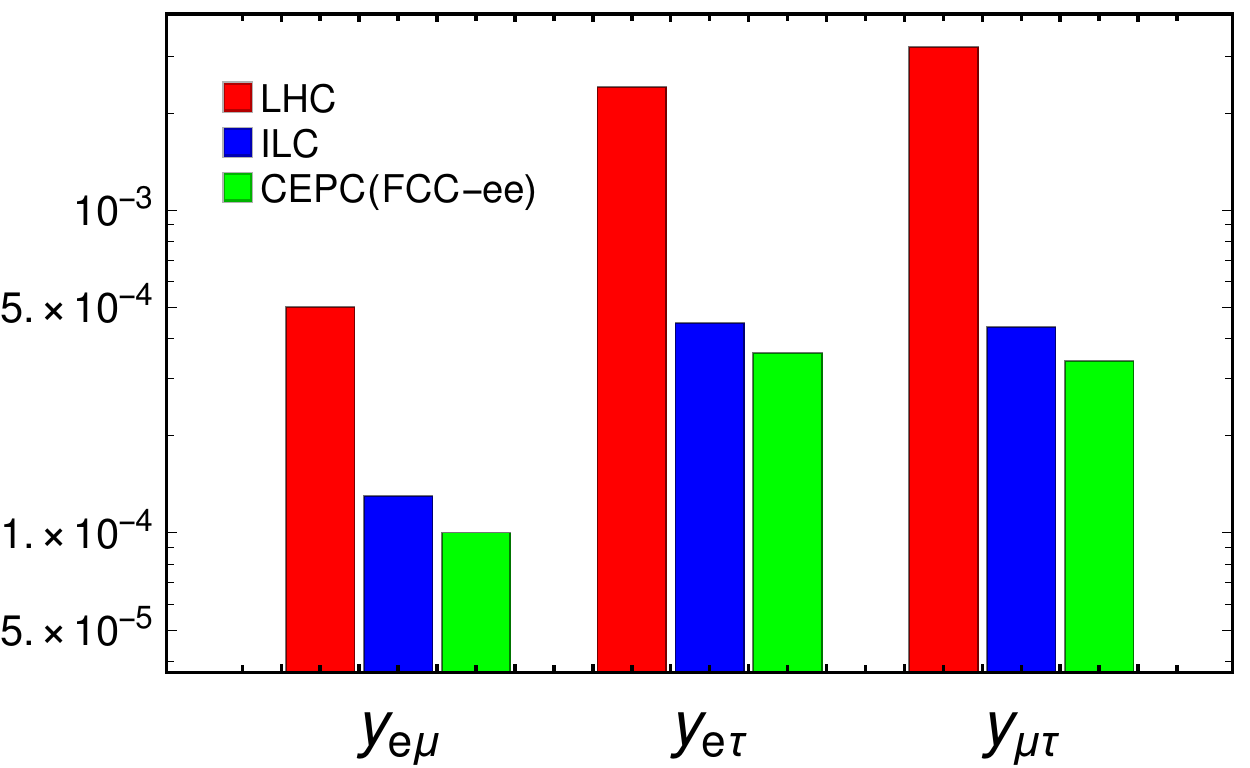}
\caption{The upper bounds at 95\% CL on the three CLFV Higgs decay rates (left) and the corresponding 
CLFV couplings (right) given by the LHC \cite{CMS:2016qvi,Khachatryan:2016rke} (red), compared to 
the corresponding bounds expected to be given by the CEPC, the FCC-ee (green) and the ILC (blue). 
See text for details.}\label{Higgscouplingbound}
\end{figure}

The upper bounds on $\mathcal{B}(H\to e^\pm\mu^\mp)$ expected to be given by the CEPC, 
the FCC-ee and the ILC are displayed in FIG. \ref{Higgscouplingbound}, together with the 
present upper bound $\mathcal{B}(H\to e^\pm\mu^\mp) < 0.035\%$ at 95\% CL reported by the 
CMS collaboration \cite{Khachatryan:2016rke}. We find that the three future lepton colliders are 
expected to improve the precision by about 30 times compared to the present CMS measurement, 
and also by one order of magnitude compared to the expected HL-LHC upper bound 
$\mathcal{B}(H\to e^\pm\mu^\mp) < \mathcal{O}(0.02)\%$ \cite{Banerjee:2016foh}.

\subsection{$H\to e^\pm\tau^\mp$}

To reconstruct the signal events of $e^+e^-\to HZ\to e^\pm\tau^\mp Z$, we again first select the events 
with two jets, one electron and one muon in their final states. In each event, the $Z$ boson is 
reconstructed using the two jets as in the $H\to e^\pm\mu^\mp$ case, the tau lepton is reconstructed from 
the muon and the missing energy, and the Higgs boson is finally reconstructed from the electron and the
tau lepton. Our tau reconstruction method has an efficiency that will allow us to 
estimate the CEPC sensitivity to $\mathcal{B}(H\to e^\pm\tau^\mp)$, since the 
branching ratio of a tau lepton decaying into a muon and two neutrinos $\mathcal{B}(\tau\to\mu\nu\bar{\nu})$ 
is nearly 20\%. Our method also greatly suppresses the background, {\it e.g.}, if we choose to reconstruct 
the tau lepton from electron and missing energy, then the processes $e^+e^-\to e^+e^-q\bar{q}$ 
would give a large background. The above event selection method requires the following 
cuts: 66 GeV $< m_{jj} <$ 94 GeV, $m_{\mu E_M} < 4$ GeV and 121 GeV $< m_{e\tau} < 130$ 
GeV, where $m_{\mu E_M}$ is the invariant mass of the muon and missing energy, and $m_{e\tau}$ 
is the invariant mass of the electron, muon and missing energy. We further set a cut on
the electron pseudorapidity of $|\eta_e|<2$ to suppress the background arising from the SZ processes.
The cuts are summarized in TABLE \ref{tb:Higgsetau_emu}, where we give the numbers of 
background events in different categories and the numbers of signal events at the CEPC after the cuts. 
We find that 456 out of the 10000 generated signal events survive the event selection, and thus the 
signal detection efficiency is $\epsilon$ = 4.56\%. As for the background, five events are expected after 
the event selection, so the upper limit on the number of signal events is $N_{95}$ = 5.51 at 95\% CL. 
Then, employing \eqref{eq:upperlimit} and recalling that the number of the Higgs bosons to be
produced by the CPEC is $N_{H}$ = 1.05 $\times 10^6$, we find the upper bound on the 
$H\to e^\pm\tau^\mp$ branching ratio as
\begin{equation}
\mathcal{B}(H\to e^\pm\tau^\mp) < 1.6\times10^{-4}\ \text{at 95\% CL},
\end{equation}
which also applies to the FCC-ee. For the ILC, we find about 1.9 background events by scaling the 
SZ$\_$ll ($e^+e^-\tau^+\tau^-$) event numbers, and hence $N_{95}$ = 4.2 and the upper bound on 
the branching ratio is $2.4\times10^{-4}$ at 95\% CL. These results and the present upper bound 
$\mathcal{B}(H\to e^\pm\tau^\mp) < 0.69\%$ at 95\% CL reported by the CMS collaboration 
\cite{Khachatryan:2016rke} (the ATLAS bound is 1.04\% \cite{Aad:2016blu}) are displayed together 
in FIG. \ref{Higgscouplingbound}. We find that the CEPC or the FCC-ee (the ILC) is expected to 
improve the sensitivity to $\mathcal{B}(H\to e^\pm\tau^\mp)$ by about 40 (30) times compared to 
the present CMS measurement, and also by one to two orders compared to the expected HL-LHC 
upper bound $\mathcal{B}(H\to e^\pm\tau^\mp) < \mathcal{O}(0.5)\%$ \cite{Banerjee:2016foh}.

\begin{table}[!htbh]
  \centering
  \begin{tabular}[t]{|l|c|c|c|c|c|c|c|c|}\hline
 \parbox[0pt][2em][c]{0cm}{} \textbf{Cut} & SZ$\_$ll & SZ$\_$ql  & ZZ$\_$ll & ZZ$\_$ql  & ZZ$\_$qq & SW$\_$ql & WW$\_$ql & \textbf{Signal}  \\ \hline
 \parbox[0pt][2em][c]{0cm}{} $N_{e,\mu}$=1, $N_j$=2 & 5684 & 1248 & 1464 & 16504 & 1945 & 1063 & 1657 & 868 \\ \hline
 \parbox[0pt][2em][c]{0cm}{} 66 GeV $< m_{jj} <$ 94 GeV & 1119 & 290 & 448 & 11828 & 417 & 305 & 641 & 693  \\ \hline
 \parbox[0pt][2em][c]{0cm}{} $m_{\mu E_M} < 4$ GeV & 423 &95 & 41 & 1892 & 47 & 7 & 0 & 530 \\ \hline
 \parbox[0pt][2em][c]{0cm}{} 121 GeV $< m_{e\tau} <$ 130 GeV & 9 & 1  & 0 & 0 & 0 & 0 & 0 & 479  \\ \hline
 \parbox[0pt][2em][c]{0cm}{} $|\eta_{e}| < 2$ & 5 & 0 & 0  & 0 & 0 & 0 & 0 & 456  \\ \hline
  \end{tabular}
  \caption{The numbers of background events in different categories and signal events surviving the cuts in the analysis of $H\to e^\pm\tau^\mp$ at the CEPC. See text for details.}\label{tb:Higgsetau_emu}
\end{table}

\subsection{$H\to\mu^\pm\tau^\mp$}

To reconstruct the signal events of $e^+e^-\to HZ\to\mu^\pm\tau^\mp Z$, we still select the events 
containing one electron, one muon and two jets in their final states. In each event, the $Z$ boson 
is reconstructed from the two jets, the tau lepton is reconstructed from the electron and the missing energy, 
and the Higgs boson is reconstructed from the muon and the reconstructed tau lepton. Analogous to the 
$H\to e^\pm\tau^\mp$ case, we do not consider other ways to reconstruct the tau lepton. We employ 
the following cuts: 60 GeV $< m_{jj} <$ 100 GeV and $m_{e E_M} < 5$ GeV and 
120 GeV $< m_{\mu\tau} < 130$ GeV, where $m_{e E_M}$ is the invariant mass of the electron and 
missing energy, and $m_{\mu\tau}$ is the invariant mass of the electron, muon and missing energy.
The cuts are listed in TABLE \ref{tb:Higgsmutau_emu}, where we also show the numbers of background 
events in different categories and the numbers of signal events after the cuts. We find that 522 out 
of the 10000 generated signal events survive the event selection, and thus the signal detection 
efficiency is $\epsilon$ = 5.22\%. As for the background, five events are expected after the event 
selection, so the upper limit on the number of signal events is $N_{95}$ = 5.51 at 95\% CL. Finally, 
we obtain the upper bound on the $H\to\mu^\pm\tau^\mp$ branching ratio given by the CEPC and 
also the FCC-ee, 
\begin{equation}
\mathcal{B}(H\to\mu^\pm\tau^\mp) < 1.4\times10^{-4}\ \text{at 95\% CL.}
\end{equation}
For the ILC, we find about 2.6 background events by scaling the ZZ$\_$ll ($\mu^+\mu^-\tau^+\tau^-$ 
and $\tau^+\tau^-\tau^+\tau^-$) and ZZ$\_$ql ($jj\tau^+\tau^-$) event numbers, and hence $N_{95}$ = 
4.6 and the upper bound on the branching ratio is $2.3\times10^{-4}$ at 95\% CL. These results and 
the present upper bound $\mathcal{B}(H\to\mu^\pm\tau^\mp) < 1.20\%$ at 95\% CL reported by the 
CMS collaboration \cite{CMS:2016qvi} (the ATLAS bound is 1.43\% \cite{Aad:2016blu}) are displayed 
together in FIG. \ref{Higgscouplingbound}. We find that any of the future lepton colliders is expected 
to improve the sensitivity to $\mathcal{B}(H\to\mu^\pm\tau^\mp)$ by nearly two orders compared to 
the present CMS measurement, and also by one to two orders compared to the expected HL-LHC 
upper bound $\mathcal{B}(H\to\mu^\pm\tau^\mp) < \mathcal{O}(0.5)\%$ \cite{Banerjee:2016foh}.

\begin{table}[!htbh]
  \centering
  \begin{tabular}[t]{|l|c|c|c|c|c|c|c|c|}\hline
 \parbox[0pt][2em][c]{0cm}{} \textbf{Cut} & SZ$\_$ll & SZ$\_$ql  & ZZ$\_$ll & ZZ$\_$ql  & ZZ$\_$qq & SW$\_$ql & WW$\_$ql & \textbf{Signal}  \\ \hline
 \parbox[0pt][2em][c]{0cm}{} $N_{e,\mu}$=1, $N_j$=2 & 5684 & 1248 & 1464 & 16504 & 1945 & 1063 & 1657 & 856 \\ \hline
 \parbox[0pt][2em][c]{0cm}{} 60 GeV $< m_{jj} <$ 100 GeV & 1578 & 428 & 606 & 13504 & 678 & 454 & 882 & 736  \\ \hline
 \parbox[0pt][2em][c]{0cm}{} $m_{e E_M} < 5$ GeV & 26 & 16 & 84 & 2706 & 54 & 0 & 48 & 583 \\ \hline
 \parbox[0pt][2em][c]{0cm}{} 120 GeV $< m_{\mu\tau} <$ 130 GeV & 0 & 0  & 2 & 3 & 0 & 0 & 0 & 522  \\ \hline
  \end{tabular}
  \caption{The numbers of background events in different categories and signal events surviving the cuts in the analysis of $H\to\mu^\pm\tau^\mp$ at the CEPC. See text for details.}\label{tb:Higgsmutau_emu}
\end{table}

According to another study of $H\to\mu^\pm\tau^\mp$ at the ILC \cite{Chakraborty:2016gff}, the upper bound 
on the branching ratio is given as $\mathcal{B}(H\to \mu^\pm\tau^\mp) < 2.9\times10^{-5}$ at 95\% CL if 90\% 
signals survive the event selection. This bound is more stringent than that obtained in this work. A possible 
reason is that we only considered the cleanest method to reconstruct the tau lepton, as described previously, 
which makes our evaluated upper bound very conservative. It is also 
necessary to point out that \cite{Chakraborty:2016gff} has assumed a tau lepton reconstruction efficiency as 
high as 70\%, muon and jet detection efficiencies as high as 100\% \cite{Kawada:2015wea}. We also suspect 
that only the $e^+e^-\to q\bar{q}\mu^\pm\tau^\mp\bar{\nu}\nu$ processes are considered as possible background 
sources in \cite{Chakraborty:2016gff} is too optimistic. The potential of the ILC to search for the $H\to\mu^\pm\tau^\mp$ 
channel has also been studied in \cite{Chakraborty:2017tyb}, where the $Z$ bosons are reconstructed using lepton 
pairs. There it is discussed that a signal with 3$\sigma$ statistical significance at the ILC with an integrated luminosity 
of 1 ab$^{-1}$ requires $H\to\mu^\pm\tau^\mp$ to have a branching ratio larger than $4.09\times 10^{-3}$.

\section{Constraints on theory parameters}\label{sec:theory}

The Lagrangian for a CLFV Higgs decay is given by 
\begin{equation}
\mathcal{L}^{H\to\ell\ell'} \ni -Y_{\ell\ell'}\bar{\ell}_LH\ell'_R - Y_{\ell'\ell}\bar{\ell}^\prime_LH\ell_R + \textit{h.c.},
\end{equation}
with $\ell\neq\ell'$. The decay width of $H\to\ell^\pm\ell^{\prime\mp}$ is then calculated to be
\begin{equation}\begin{split}
\Gamma(H\to\ell^\pm\ell^{\prime\mp}) = {m_H\over 8\pi}\left|y_{\ell\ell'}\right|^2
\end{split}\end{equation}
in the zero lepton mass limit, where $y_{\ell\ell'}$ is defined by $y_{\ell\ell'}\equiv\sqrt{|Y_{\ell\ell'}|^2+|Y_{\ell'\ell}|^2}$. 
Assuming that new physics only enters via the $H\ell\ell'$ coupling, the $H\to\ell^\pm\ell^{\prime\mp}$ branching ratio is 
given by 
\begin{equation}\label{eq:br}
\mathcal{B}(H\to\ell^\pm\ell^{\prime\mp}) = {\Gamma(H\to\ell^\pm\ell^{\prime\mp}) \over \Gamma(H\to\ell^\pm\ell^{\prime\mp}) + \Gamma_\text{SM}},
\end{equation}
where the SM Higgs boson decay width is $\Gamma_\text{SM}$ = 4.1 MeV \cite{Denner:2011mq}.
This leads to the upper bounds on the CLFV Higgs couplings expected to be given by the three lepton colliders, 
\begin{align}
\text{CEPC(FCC-ee):}&~&y_{e\mu} < 1.0\times 10^{-4},\ y_{e\tau} < 3.6\times 10^{-4},\ y_{\mu\tau} < 3.4\times 10^{-4}\ \text{at 95\% CL,}\nonumber \\
\text{ILC:}&~&y_{e\mu} < 1.3\times 10^{-4},\ y_{e\tau} < 4.5\times 10^{-4},\ y_{\mu\tau} < 4.3\times 10^{-4} \ \text{at 95\% CL.}
\end{align}
As a comparison, we also list the current experimental bounds on the CLFV Higgs couplings,
\begin{equation}
y_{e\mu} < 0.5\times 10^{-3},\qquad
y_{e\tau} < 2.4\times 10^{-3},\qquad y_{\mu\tau} < 3.2\times 10^{-3} \ \text{at 95\% CL,}
\end{equation}
which are obtained from the LHC bounds on the corresponding branching ratios \cite{CMS:2016qvi,Khachatryan:2016rke}.
All these upper bounds on the $H\ell\ell'$ couplings are displayed in FIG. \ref{Higgscouplingbound}.

\subsection*{\underline{Constraints on the SMEFT}}

We also consider the constraints on the new physics cut-off scale $\Lambda$ implicated by the 
improved bounds on the CLFV Higgs decay rates in the SMEFT 
\cite{Buchmuller:1985jz,Grzadkowski:2010es}, which contains higher-dimension 
operators invariant under the SM gauge transformations. The dimension-six operators 
$H^\dagger H\bar{f}'_iHf''_j$ result in the fermions coupling to the Higgs vacuum expectation 
$v$ differently from to the Higgs boson after the spontaneous symmetry breaking, and the off-diagonal entries of 
the $Hf_if_j$ coupling matrices are proportional to ${v^2\over\sqrt{2}\Lambda^2}$ 
\cite{Harnik:2012pb,Alonso:2013hga,Pruna:2014asa,Pruna:2015jhf}, namely 
\begin{equation}\label{yukawa}
Y_{ij} = {v^2\over\sqrt{2}\Lambda^2}C_{ij},
\end{equation}
with $f_{i,j}$ the mass eigenstates and $i\neq j$. Assuming $C_{ij}\sim 1$, the expected CEPC (ILC) constraint on
the $H\to e^\pm\mu^\mp$ branching ratio will give the most stringent lower bound on $\Lambda$, 
$\Lambda \gtrsim 25$ (22) TeV. However, the order of $C_{ij}$ depends crucially on flavor structures beyond the 
SM. If we adopt the Cheng-Sher ansatz $C_{ij}\sim \sqrt{m_{i}m_{j}}/v$ \cite{Cheng:1987rs}, the 
$H\to \mu^\pm\tau^\mp$ channel will set the most stringent lower bound on $\Lambda$, which reads 
$\Lambda \gtrsim 0.6$ (0.5) TeV.

\subsection*{\underline{Constraints on the type III 2HDM}}

Although one Higgs field is enough for electroweak symmetry breaking and giving masses to gauge bosons 
and fermions as in the SM, there are several motivations for introduction of two Higgs doublets \cite{Lee:1973iz}, 
including requirement of supersymmetry \cite{Haber:1984rc}, axion models \cite{Kim:1986ax} and baryogenesis 
\cite{Turok:1990zg} (see {\it e.g.} \cite{Branco:2011iw}). In 
a two-Higgs-doublet-model (2HDM), couplings of the Higgs boson with other particles are 
modified compared to the SM. Especially, the type III 2HDM naturally introduces tree-level CLFV Higgs couplings. 
In the type III 2HDM, two doublets $\Phi_1 = 1/\sqrt{2}( ..., v_1+\rho_1 + ...)^\text{T}$ and 
$\Phi_2 = 1/\sqrt{2}( ..., v_2+\rho_2 + ...)^\text{T}$ with hypercharge +1 couple to fermions freely. We can 
rotate the scalar doublets such that the vacuum expectation is entirely in the first doublet, 
\begin{equation}
\begin{array}{c}
H_1=\Phi_1\cos\beta + \Phi_2\sin\beta,\\
H_2=\Phi_1\sin\beta - \Phi_2\cos\beta,
\end{array}
\qquad
\langle H_1\rangle = \left(\begin{array}{c} 0\\ v/\sqrt{2} \\ \end{array} \right),
\qquad
\langle H_2\rangle = \left(\begin{array}{c} 0\\ 0 \\ \end{array} \right),
\end{equation}
with $v=\sqrt{v_1^2+v_2^2}$ and $\tan\beta = v_2/v_1$.
While the mass eigenstates are given by another rotation, 
\begin{equation}
H = \rho_1\sin\alpha - \rho_2\cos\alpha,\qquad
H' = - \rho_1\cos\alpha - \rho_2\sin\alpha,
\end{equation}
and equivalently 
\begin{equation}
\begin{split}
H=&\ H_1^0\sin(\alpha-\beta) + H_2^0\cos(\alpha-\beta),\\
H'=&-H_1^0\cos(\alpha-\beta) - H_2^0\sin(\alpha - \beta).
\end{split}
\end{equation}
Diagonalizing the mass matrix automatically diagonalize the $H_1^0$ coupling matrix, so the CLHV 
vertices only come from $H_2^0$. Therefore, the CLFV couplings with $H$ is given by 
\begin{equation}
\mathcal{L}^{H\to\ell\ell'} \ni -\cos(\alpha-\beta)\xi_{\ell\ell'}\bar{\ell}_LH\ell'_R - \cos(\alpha-\beta)\xi_{\ell'\ell}\bar{\ell}^\prime_LH\ell_R + \textit{h.c.}.
\end{equation}
Under the Cheng-Sher ansatz \cite{Cheng:1987rs}, we define $\xi_{ij} = \lambda_{ij}\sqrt{2m_im_j}/v$, 
where $\lambda_{ij}$ are of order one. On the other hand, the flavor conserving $H$ couplings receive 
contributions from both $H_1^0$ and $H_2^0$. For example, the expression for the $Hb\bar{b}$ coupling 
is given by $y_{b}\sin(\alpha-\beta)+\xi_{b}\cos(\alpha-\beta)$ with $y_b=m_b/v$, and we further write 
$\xi_b=\lambda_b m_b/v$ and expect $\lambda_b$ to be of order one. 

\begin{figure}[ht!]
\centering
\includegraphics[width=73mm]{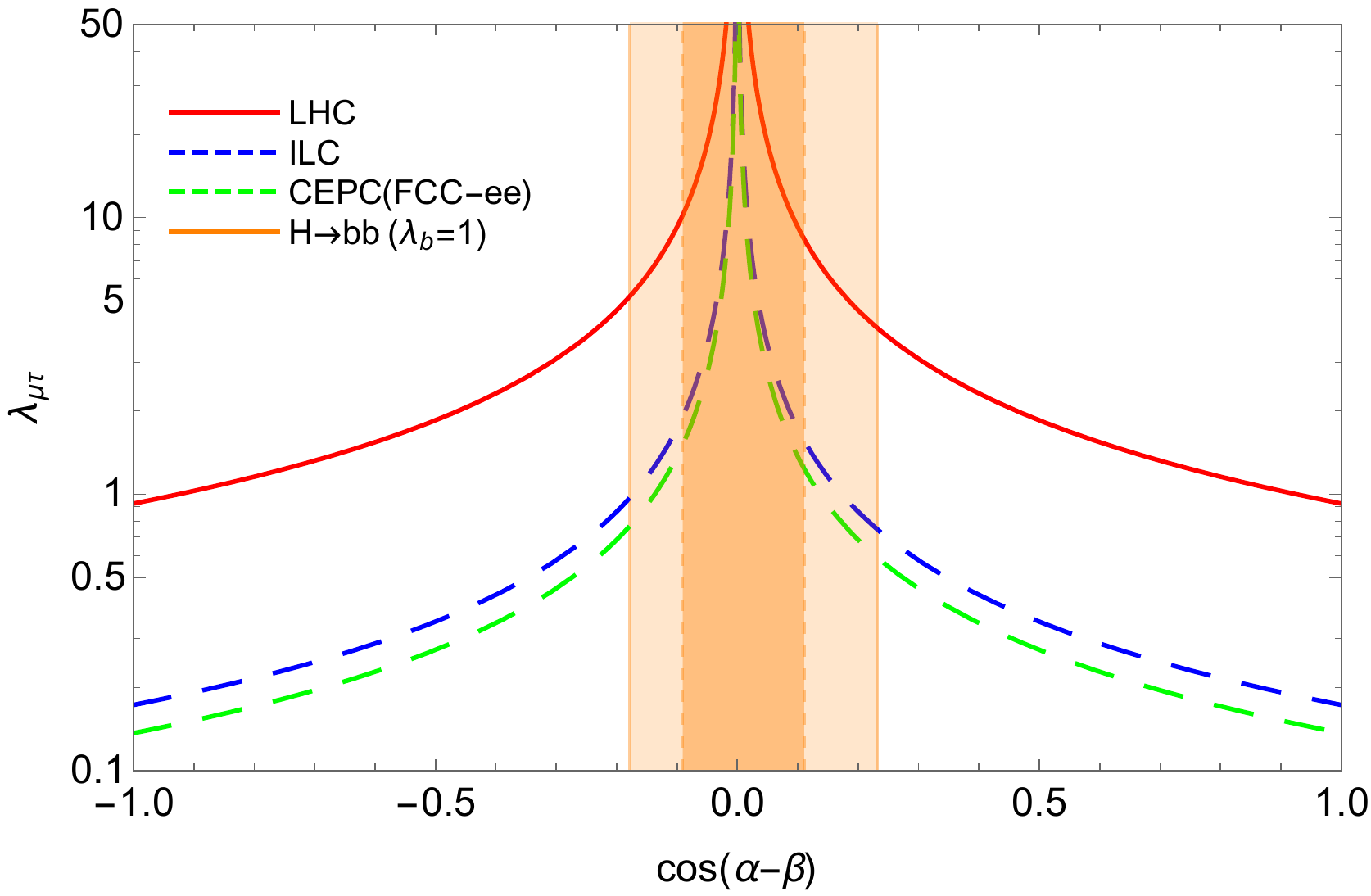}
  \hspace{0.2in}
\includegraphics[width=73mm]{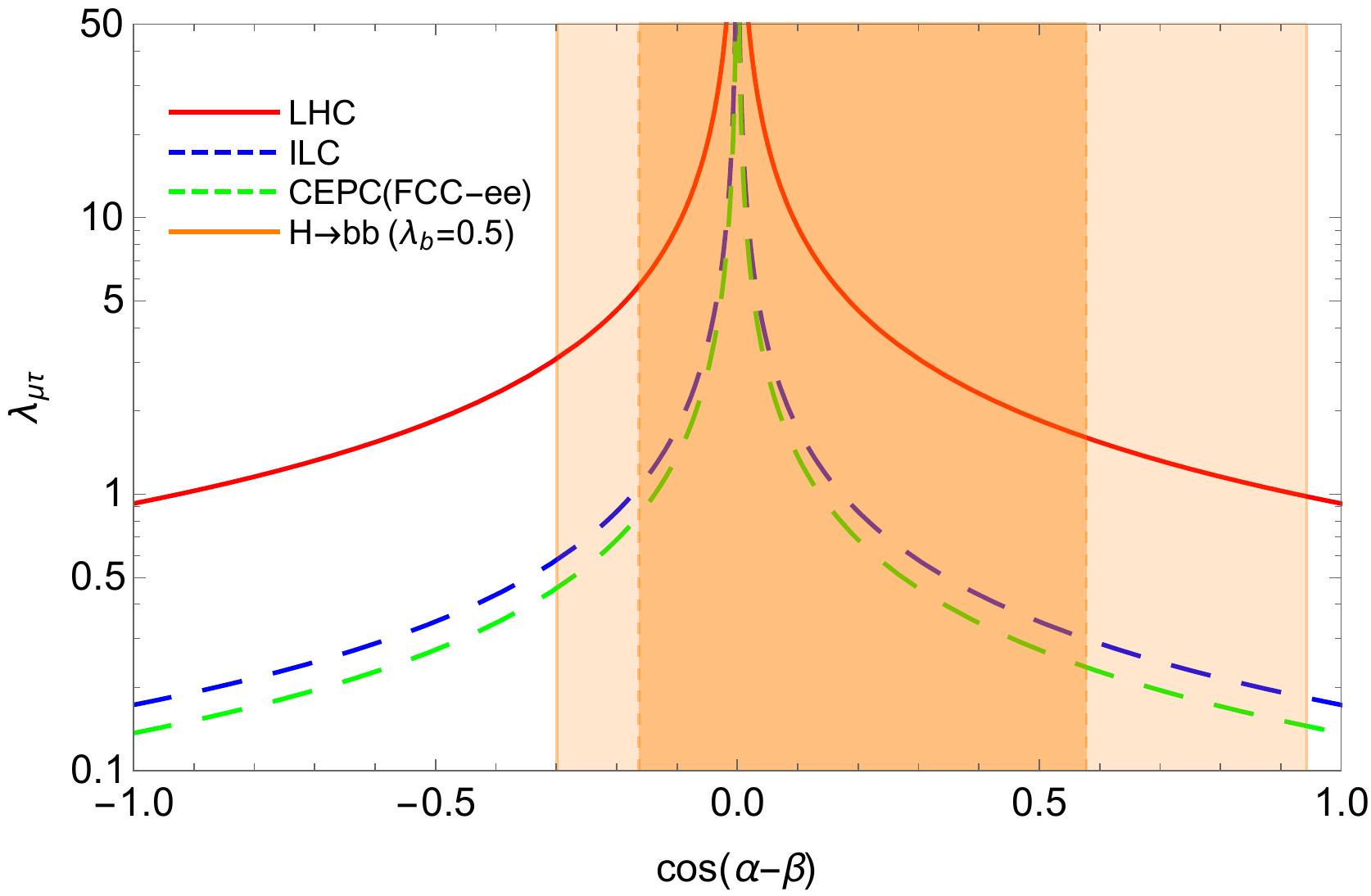}
\caption{The red (solid), blue (dashed) and green (dashed) curves represent the constraints at 95\% 
CL on the $\cos(\alpha-\beta)$-$\lambda_{\mu\tau}$ plain (the regions above the curves are excluded) 
set by the the LHC \cite{CMS:2016qvi}, the ILC and the CEPC (FCC-ee) upper bounds on 
$\mathcal{B}(H\to\mu^\pm\tau^\mp)$, respectively. The orange (light orange) contours are the 
corresponding 1(2)-sigma allowed ranges of $\cos(\alpha-\beta)$ by the ATLAS measurement of the 
$H\to b\bar{b}$ signal strength \cite{Aaboud:2018zhk} with $\lambda_b$ = 1 (left) and 0.5 (right). 
See text for details.}
\label{2HDMbound}
\end{figure}

According to the current measurements, no obvious deviation of Higgs couplings from the SM  
has been found \cite{Mariotti:2016owy}.  This is further confirmed by the recent observation of the 
$H\to b\bar{b}$ channel by the CMS \cite{Sirunyan:2018kst} and the ATLAS \cite{Aaboud:2018zhk} 
collaborations. It indicates that the parameters $\lambda_{i(j)}$ and $\alpha-\beta$ are strictly constrained, 
which also makes \eqref{eq:br} approximately valid. Here we study how the bounds 
on the CLFV Higgs decay rates together with the $Hb\bar{b}$ coupling set constraints on the relevant 
parameters $\lambda_{\ell\ell'}$ and $\cos(\alpha-\beta)$ in the 2HDM. The $H\to\mu^\pm\tau^\mp$ 
channel is taken as an example. From the ATLAS measurement of the $H\to b\bar{b}$ signal strength 
\cite{Aaboud:2018zhk}, it can be extracted that the ratio of the $Hb\bar{b}$ coupling to the SM expectation 
is $1.005\pm0.10$. We choose the order-one $\lambda_b$ to be 1 and 0.5 as two benchmarks, and 
display the corresponding 1- and 2-sigma allowed ranges of $\cos(\alpha-\beta)$ in the two panels of 
FIG. \ref{2HDMbound} by the orange and light orange contours. In FIG. \ref{2HDMbound}, we also show 
on the $\cos(\alpha-\beta)$-$\lambda_{\mu\tau}$ plain the constraints (the regions above the curves 
are excluded) set by the upper bounds on $\mathcal{B}(H\to\mu^\pm\tau^\mp)$. It is observed that 
$\cos(\alpha-\beta)$ still has a large living space especially in the $\lambda_b$ = 0.5 case,  and that 
in the region when $|\cos(\alpha-\beta)|$ is large, the upper bounds on $\mathcal{B}(H\to\mu^\pm\tau^\mp)$ 
given by the the future lepton colliders restrict $\lambda_{\mu\tau}$ to be smaller than $\mathcal{O}(0.1)$.

\subsection*{\underline{Constraints on RS models}}

In RS models \cite{Randall:1999ee,Randall:1999vf} in which the fermions are allowed to propagate in 
the extra dimension, the large fermion mass hierarchies and the tiny neutrino masses can be explained
\cite{Gherghetta:2000qt,Grossman:1999ra}. In order to generate the observed structure in the lepton 
sector, which means the hierarchies between charged lepton masses, the neutrino masses with a similar size
and the large neutrino mixing angles, one can, in the assumption of Dirac neutrinos, set same or similar profiles 
in the fifth dimension for the lepton doublets and neutrino singlets and set different profiles for the charged 
lepton singlets (see \textit{e.g.} \cite{Frank:2014aca}). In such a case, we find that the coefficient in 
\eqref{yukawa} $C_{ij}\sim m_\tau/v$ (see Section 4.1 of \cite{Casagrande:2008hr}). Therefore, in such models, the most stringent 
lower bound on $\Lambda$, or the famous Kaluza-Klein scale $M_\text{KK}$ \cite{Kaluza:1921tu,Klein:1926tv}, 
is set by the expected CEPC (ILC) constraint on the $H\to \mu^\pm\tau^\mp$ branching ratio as 
$\Lambda\gtrsim 2.5$ (2.2) TeV. Since the masses of the lightest Kaluza-Klein particles are approximately 
2.45$M_\text{KK}$, nondiscovery of $H\to \mu^\pm\tau^\mp$ at the CEPC (ILC) will excluded Kaluza-Klein 
particles with masses smaller than 6.1 (5.4) TeV.

\subsection*{\underline{Constraints on models with heavy neutrinos}}

Lepton flavor violation may originate from heavy neutrinos at one-loop level (see {\it e.g.} \cite{Ilakovac:1994kj}). 
Taking the Inverse Seesaw Model as an example with the right-handed neutrino masses $M_R$ close 
to the TeV scale, we have approximately the off-diagonal charged lepton Yukawa couplings 
\begin{equation}\label{heavynu}
Y_{ij} \approx {g\over64\pi^2}{m_i\over m_W}\left[{m_H^2\over m_R^2} \left(r({m_W^2\over m_H^2}) 
+ \log({m_W^2\over m_H^2})\right)(Y_\nu Y_\nu^\dagger)_{ij} -{3v^2\over 2M_R^2}(Y_\nu Y_\nu^\dagger Y_\nu Y_\nu^\dagger)_{ij} \right]
\end{equation}
according to (25) of \cite{Arganda:2017vdb}, with $Y_\nu$ the neutrino Yukawa 
coupling matrix, $g$ the $SU(2)$ gauge coupling constant, $m_i$ the mass of the $i$th generation charged lepton, 
$m_W$ the $W$ boson mass and $m_H$ the Higgs boson mass. The function $r(\lambda)$ is given by (26) of 
\cite{Arganda:2017vdb}. If we assume a benchmark neutrino Yukawa coupling matrix following \cite{Arganda:2017vdb}, 
$Y_\nu = \{ (0.1, 0, 0), (0, 1, 0), (0, 1, 0.014)\}$, a rough calculation indicates that the lower bound on 
$M_R$ set by the expected measurements of the CLFV Higgs decay channels at the three future 
lepton colliders will be $M_R \gtrsim$ 0.3 GeV. Since such a small right-handed neutrino mass would
not satisfy the perturbation condition, we conclude that the expected improved bounds on the CLFV Higgs 
decay rates put no constraint on the right-handed neutrino masses. The complete expression for the 
off-diagonal charged lepton Yukawa couplings, which does not rely on the expansion in inverse powers of $M_R$, 
can be found in Appendix C of \cite{Marcano:2017ucg}. As discussed in \cite{Arganda:2017vdb}, using 
\eqref{heavynu} always overestimates values of $Y_{ij}$, so using the complete formula will give an even looser 
constraint on $M_R$, which will not change our main conclusion.

\section{Summary}\label{sec:summary}

The future $e^+e^-$ colliders, the CEPC, the FCC-ee and the ILC,  as Higgs factories, are ideal machines 
for precise studies of Higgs properties. In this paper, we evaluate the potential of the three lepton colliders
for searching for the CLFV Higgs decays. We find that the expected upper bounds given by the CEPC or the 
FCC-ee (the ILC) on the branching ratios of $H\to e^\pm\mu^\mp$, $e^\pm\tau^\mp$ and $\mu^\pm\tau^\mp$ 
are $1.2\ (2.1)\times10^{-5}$, 
$1.6\ (2.4)\times10^{-4}$ and $1.4\ (2.3)\times10^{-4}$ at 95\% CL, respectively. The resulting constraints on certain theory
parameters are also given, including the CLFV Higgs couplings, the relevant parameters in the type-III 2HDM, 
and the cut-off scales in the SMEFT and in RS models.

\section*{Acknowledgement}
The authors are grateful to Simon Brass, Cheng Chen, Wolfgang Kilian, Gang Li, Yichen Li, Manqi Ruan, 
Lei Wang, Jue Zhang and Zhijie Zhao for enlightening discussions, and to Keri Vos for English language 
revision. This work was supported in part by by DFG Forschergruppe FOR 1873 "Quark Flavour Physics 
and Effective Field Theories", by National Natural Science Foundation of China under Grant No. 11235005, 
11347027, 11375208, 11475180, 11505083, 11521505, 11575005, 11621131001 and U1732101. In addition, 
Qin Qin thanks Xun Luo for brightening up his life, and here he proposes to her, "will you marry me?"


\begin{thebibliography}{99}

%\cite{Aad:2012tfa}
\bibitem{Aad:2012tfa} 
  G.~Aad {\it et al.} [ATLAS Collaboration],
  %``Observation of a new particle in the search for the Standard Model Higgs boson with the ATLAS detector at the LHC,''
  Phys.\ Lett.\ B {\bf 716}, 1 (2012)
  doi:10.1016/j.physletb.2012.08.020
  [arXiv:1207.7214 [hep-ex]].
  %%CITATION = doi:10.1016/j.physletb.2012.08.020;%%
  %8807 citations counted in INSPIRE as of 09 Sep 2018


%\cite{Chatrchyan:2012xdj}
\bibitem{Chatrchyan:2012xdj} 
  S.~Chatrchyan {\it et al.} [CMS Collaboration],
  %``Observation of a new boson at a mass of 125 GeV with the CMS experiment at the LHC,''
  Phys.\ Lett.\ B {\bf 716}, 30 (2012)
  doi:10.1016/j.physletb.2012.08.021
  [arXiv:1207.7235 [hep-ex]].
  %%CITATION = doi:10.1016/j.physletb.2012.08.021;%%
  %8599 citations counted in INSPIRE as of 09 Sep 2018


%\cite{Chatrchyan:2013lba}
\bibitem{Chatrchyan:2013lba} 
  S.~Chatrchyan {\it et al.} [CMS Collaboration],
  %``Observation of a new boson with mass near 125 GeV in pp collisions at $\sqrt{s}$ = 7 and 8 TeV,''
  JHEP {\bf 1306}, 081 (2013)
  doi:10.1007/JHEP06(2013)081
  [arXiv:1303.4571 [hep-ex]].
  %%CITATION = doi:10.1007/JHEP06(2013)081;%%
  %671 citations counted in INSPIRE as of 09 Sep 2018


%\cite{Aad:2013wqa}
\bibitem{Aad:2013wqa} 
  G.~Aad {\it et al.} [ATLAS Collaboration],
  %``Measurements of Higgs boson production and couplings in diboson final states with the ATLAS detector at the LHC,''
  Phys.\ Lett.\ B {\bf 726}, 88 (2013)
  Erratum: [Phys.\ Lett.\ B {\bf 734}, 406 (2014)]
  doi:10.1016/j.physletb.2014.05.011, 10.1016/j.physletb.2013.08.010
  [arXiv:1307.1427 [hep-ex]].
  %%CITATION = doi:10.1016/j.physletb.2014.05.011, 10.1016/j.physletb.2013.08.010;%%
  %633 citations counted in INSPIRE as of 09 Sep 2018


%\cite{Aad:2013xqa}
\bibitem{Aad:2013xqa} 
  G.~Aad {\it et al.} [ATLAS Collaboration],
  %``Evidence for the spin-0 nature of the Higgs boson using ATLAS data,''
  Phys.\ Lett.\ B {\bf 726}, 120 (2013)
  doi:10.1016/j.physletb.2013.08.026
  [arXiv:1307.1432 [hep-ex]].
  %%CITATION = doi:10.1016/j.physletb.2013.08.026;%%
  %625 citations counted in INSPIRE as of 09 Sep 2018


%\cite{Khachatryan:2014jba}
\bibitem{Khachatryan:2014jba} 
  V.~Khachatryan {\it et al.} [CMS Collaboration],
  %``Precise determination of the mass of the Higgs boson and tests of compatibility of its couplings with the standard model predictions using proton collisions at 7 and 8 $\,\text {TeV}$,''
  Eur.\ Phys.\ J.\ C {\bf 75}, no. 5, 212 (2015)
  doi:10.1140/epjc/s10052-015-3351-7
  [arXiv:1412.8662 [hep-ex]].
  %%CITATION = doi:10.1140/epjc/s10052-015-3351-7;%%
  %720 citations counted in INSPIRE as of 09 Sep 2018


%\cite{Khachatryan:2014kca}
\bibitem{Khachatryan:2014kca} 
  V.~Khachatryan {\it et al.} [CMS Collaboration],
  %``Constraints on the spin-parity and anomalous HVV couplings of the Higgs boson in proton collisions at 7 and 8 TeV,''
  Phys.\ Rev.\ D {\bf 92}, no. 1, 012004 (2015)
  doi:10.1103/PhysRevD.92.012004
  [arXiv:1411.3441 [hep-ex]].
  %%CITATION = doi:10.1103/PhysRevD.92.012004;%%
  %325 citations counted in INSPIRE as of 09 Sep 2018


%\cite{Aad:2015zhl}
\bibitem{Aad:2015zhl} 
  G.~Aad {\it et al.} [ATLAS and CMS Collaborations],
  %``Combined Measurement of the Higgs Boson Mass in $pp$ Collisions at $\sqrt{s}=7$ and 8 TeV with the ATLAS and CMS Experiments,''
  Phys.\ Rev.\ Lett.\  {\bf 114}, 191803 (2015)
  doi:10.1103/PhysRevLett.114.191803
  [arXiv:1503.07589 [hep-ex]].
  %%CITATION = doi:10.1103/PhysRevLett.114.191803;%%
  %1178 citations counted in INSPIRE as of 09 Sep 2018


%\cite{CEPCStudyGroup:2018rmc}
\bibitem{CEPCStudyGroup:2018rmc} 
  [CEPC Study Group],
  %``CEPC Conceptual Design Report,''
  arXiv:1809.00285 [physics.acc-ph].
  %%CITATION = ARXIV:1809.00285;%%


%\cite{Gomez-Ceballos:2013zzn}
\bibitem{Gomez-Ceballos:2013zzn} 
  M.~Bicer {\it et al.} [TLEP Design Study Working Group],
  %``First Look at the Physics Case of TLEP,''
  JHEP {\bf 1401}, 164 (2014)
  doi:10.1007/JHEP01(2014)164
  [arXiv:1308.6176 [hep-ex]].
  %%CITATION = doi:10.1007/JHEP01(2014)164;%%
  %445 citations counted in INSPIRE as of 09 Sep 2018


%\cite{Baer:2013cma}
\bibitem{Baer:2013cma} 
  H.~Baer {\it et al.},
  %``The International Linear Collider Technical Design Report - Volume 2: Physics,''
  arXiv:1306.6352 [hep-ph].
  %%CITATION = ARXIV:1306.6352;%%
  %648 citations counted in INSPIRE as of 09 Sep 2018


%\cite{Mo:2015mza}
\bibitem{Mo:2015mza} 
  X.~Mo, G.~Li, M.~Q.~Ruan and X.~C.~Lou,
  %``Physics cross sections and event generation of $e^+e^-$ annihilations at the CEPC,''
  Chin.\ Phys.\ C {\bf 40}, no. 3, 033001 (2016)
  doi:10.1088/1674-1137/40/3/033001
  [arXiv:1505.01008 [hep-ex]].
  %%CITATION = doi:10.1088/1674-1137/40/3/033001;%%
  %19 citations counted in INSPIRE as of 09 Sep 2018


%\cite{Chen:2016zpw}
\bibitem{Chen:2016zpw} 
  Z.~Chen, Y.~Yang, M.~Ruan, D.~Wang, G.~Li, S.~Jin and Y.~Ban,
  %``Cross Section and Higgs Mass Measurement with Higgsstrahlung at the CEPC,''
  Chin.\ Phys.\ C {\bf 41}, no. 2, 023003 (2017)
  doi:10.1088/1674-1137/41/2/023003
  [arXiv:1601.05352 [hep-ex]].
  %%CITATION = doi:10.1088/1674-1137/41/2/023003;%%
  %15 citations counted in INSPIRE as of 09 Sep 2018


%\cite{Xu:2015goa}
\bibitem{Xu:2015goa} 
  G.~Z.~Xu, G.~Li, Y.~J.~Li, K.~Y.~Liu and Y.~J.~Zhang,
  %``Interference effects on Higgs mass measurement in $e^+e^-\to H(\gamma\gamma) Z$ at CEPC,''
  Chin.\ Phys.\ C {\bf 40}, no. 3, 033101 (2016)
  doi:10.1088/1674-1137/40/3/033101
  [arXiv:1505.06981 [hep-ph]].
  %%CITATION = doi:10.1088/1674-1137/40/3/033101;%%
  %2 citations counted in INSPIRE as of 09 Sep 2018


%\cite{Gong:2016jys}
\bibitem{Gong:2016jys} 
  Y.~Gong, Z.~Li, X.~Xu, L.~L.~Yang and X.~Zhao,
  %``Mixed QCD-EW corrections for Higgs boson production at $e^+e^-$ colliders,''
  Phys.\ Rev.\ D {\bf 95}, no. 9, 093003 (2017)
  doi:10.1103/PhysRevD.95.093003
  [arXiv:1609.03955 [hep-ph]].
  %%CITATION = doi:10.1103/PhysRevD.95.093003;%%
  %8 citations counted in INSPIRE as of 09 Sep 2018


%\cite{Chen:2016prx}
\bibitem{Chen:2016prx} 
  Z.~Chen {\it et al.} [CEPC Collaboration],
  %``Higgs recoil analysis and Higgs width measurement at CEPC,''
  PoS ICHEP {\bf 2016}, 432 (2016).
  doi:10.22323/1.282.0432
  %%CITATION = doi:10.22323/1.282.0432;%%
  %1 citations counted in INSPIRE as of 09 Sep 2018


%\cite{Liu:2016zki}
\bibitem{Liu:2016zki} 
  Z.~Liu, L.~T.~Wang and H.~Zhang,
  %``Exotic decays of the 125 GeV Higgs boson at future $e^+e^-$ lepton colliders,''
  Chin.\ Phys.\ C {\bf 41}, no. 6, 063102 (2017)
  doi:10.1088/1674-1137/41/6/063102
  [arXiv:1612.09284 [hep-ph]].
  %%CITATION = doi:10.1088/1674-1137/41/6/063102;%%
  %14 citations counted in INSPIRE as of 09 Sep 2018


%\cite{Chen:2017ipx}
\bibitem{Chen:2017ipx} 
  C.~Chen, Z.~Cui, G.~Li, Q.~Li, M.~Ruan, L.~Wang and Q.~s.~Yan,
  %``$H \rightarrow e^+ e^- $ at CEPC: ISR effect with MadGraph,''
  arXiv:1705.04486 [hep-ph].
  %%CITATION = ARXIV:1705.04486;%%
  %2 citations counted in INSPIRE as of 09 Sep 2018


%\cite{Cui:2017dqp}
\bibitem{Cui:2017dqp} 
  Z.~Cui, G.~Li, Q.~Li, M.~Ruan, L.~Wang and D.~Yang,
  %``Measurement of $\mathrm{H} \to \mu^+ \mu^-$ production in association with a Z boson at the CEPC,''
  Chin.\ Phys.\ C {\bf 42}, no. 5, 053001 (2018)
  doi:10.1088/1674-1137/42/5/053001
  [arXiv:1711.06807 [hep-ex]].
  %%CITATION = doi:10.1088/1674-1137/42/5/053001;%%


%\cite{Ali:2018ifm}
\bibitem{Ali:2018ifm} 
  A.~Ali, A.~Y.~Parkhomenko, Q.~Qin and W.~Wang,
  %``Prospects of discovering stable double-heavy tetraquarks at a Tera-$Z$ factory,''
  Phys.\ Lett.\ B {\bf 782}, 412 (2018)
  doi:10.1016/j.physletb.2018.05.055
  [arXiv:1805.02535 [hep-ph]].
  %%CITATION = doi:10.1016/j.physletb.2018.05.055;%%
  %3 citations counted in INSPIRE as of 09 Sep 2018
  
%\cite{He:2017ord}
\bibitem{He:2017ord} 
  M.~He, X.~G.~He and C.~K.~Huang,
  %``Dark Photon Search at A Circular $e^+e^-$ Collider,''
  Int.\ J.\ Mod.\ Phys.\ A {\bf 32}, no. 23n24, 1750138 (2017)
  doi:10.1142/S0217751X1750138X
  [arXiv:1701.08614 [hep-ph]].
  %%CITATION = doi:10.1142/S0217751X1750138X;%%
  %10 citations counted in INSPIRE as of 09 Sep 2018


%\cite{Cao:2016qgc}
\bibitem{Cao:2016qgc} 
  Q.~H.~Cao, Y.~Li, B.~Yan, Y.~Zhang and Z.~Zhang,
  %``Probing dark particles indirectly at the CEPC,''
  Nucl.\ Phys.\ B {\bf 909}, 197 (2016)
  doi:10.1016/j.nuclphysb.2016.05.010
  [arXiv:1604.07536 [hep-ph]].
  %%CITATION = doi:10.1016/j.nuclphysb.2016.05.010;%%
  %5 citations counted in INSPIRE as of 09 Sep 2018


%\cite{Liu:2017lpo}
\bibitem{Liu:2017lpo} 
  J.~Liu, X.~P.~Wang and F.~Yu,
  %``A Tale of Two Portals: Testing Light, Hidden New Physics at Future $e^+ e^-$ Colliders,''
  JHEP {\bf 1706}, 077 (2017)
  doi:10.1007/JHEP06(2017)077
  [arXiv:1704.00730 [hep-ph]].
  %%CITATION = doi:10.1007/JHEP06(2017)077;%%
  %11 citations counted in INSPIRE as of 09 Sep 2018


%\cite{Cai:2017wdu}
\bibitem{Cai:2017wdu} 
  C.~Cai, Z.~H.~Yu and H.~H.~Zhang,
  %``CEPC Precision of Electroweak Oblique Parameters and Weakly Interacting Dark Matter: the Scalar Case,''
  Nucl.\ Phys.\ B {\bf 924}, 128 (2017)
  doi:10.1016/j.nuclphysb.2017.09.007
  [arXiv:1705.07921 [hep-ph]].
  %%CITATION = doi:10.1016/j.nuclphysb.2017.09.007;%%
  %7 citations counted in INSPIRE as of 09 Sep 2018


%\cite{Xiang:2017yfs}
\bibitem{Xiang:2017yfs} 
  Q.~F.~Xiang, X.~J.~Bi, P.~F.~Yin and Z.~H.~Yu,
  %``Exploring Fermionic Dark Matter via Higgs Boson Precision Measurements at the Circular Electron Positron Collider,''
  Phys.\ Rev.\ D {\bf 97}, no. 5, 055004 (2018)
  doi:10.1103/PhysRevD.97.055004
  [arXiv:1707.03094 [hep-ph]].
  %%CITATION = doi:10.1103/PhysRevD.97.055004;%%
  %9 citations counted in INSPIRE as of 09 Sep 2018


%\cite{Wang:2017sxx}
\bibitem{Wang:2017sxx} 
  J.~W.~Wang, X.~J.~Bi, Q.~F.~Xiang, P.~F.~Yin and Z.~H.~Yu,
  %``Exploring triplet-quadruplet fermionic dark matter at the LHC and future colliders,''
  Phys.\ Rev.\ D {\bf 97}, no. 3, 035021 (2018)
  doi:10.1103/PhysRevD.97.035021
  [arXiv:1711.05622 [hep-ph]].
  %%CITATION = doi:10.1103/PhysRevD.97.035021;%%
  %2 citations counted in INSPIRE as of 09 Sep 2018


%\cite{Antusch:2015rma}
\bibitem{Antusch:2015rma} 
  S.~Antusch and O.~Fischer,
  %``Testing sterile neutrino extensions of the Standard Model at the Circular Electron Positron Collider,''
  Int.\ J.\ Mod.\ Phys.\ A {\bf 30}, no. 23, 1544004 (2015).
  doi:10.1142/S0217751X15440042
  %%CITATION = doi:10.1142/S0217751X15440042;%%
  %14 citations counted in INSPIRE as of 09 Sep 2018


%\cite{Antusch:2017pkq}
\bibitem{Antusch:2017pkq} 
  S.~Antusch, E.~Cazzato, M.~Drewes, O.~Fischer, B.~Garbrecht, D.~Gueter and J.~Klaric,
  %``Probing Leptogenesis at Future Colliders,''
  arXiv:1710.03744 [hep-ph].
  %%CITATION = ARXIV:1710.03744;%%
  %16 citations counted in INSPIRE as of 09 Sep 2018


%\cite{Liao:2017jiz}
\bibitem{Liao:2017jiz} 
  W.~Liao and X.~H.~Wu,
  %``Signature of heavy sterile neutrinos at CEPC,''
  Phys.\ Rev.\ D {\bf 97}, no. 5, 055005 (2018)
  doi:10.1103/PhysRevD.97.055005
  [arXiv:1710.09266 [hep-ph]].
  %%CITATION = doi:10.1103/PhysRevD.97.055005;%%
  %3 citations counted in INSPIRE as of 09 Sep 2018


%\cite{Banerjee:2015gca}
\bibitem{Banerjee:2015gca} 
  S.~Banerjee, P.~S.~B.~Dev, A.~Ibarra, T.~Mandal and M.~Mitra,
  %``Prospects of Heavy Neutrino Searches at Future Lepton Colliders,''
  Phys.\ Rev.\ D {\bf 92}, 075002 (2015)
  doi:10.1103/PhysRevD.92.075002
  [arXiv:1503.05491 [hep-ph]].
  %%CITATION = doi:10.1103/PhysRevD.92.075002;%%
  %71 citations counted in INSPIRE as of 09 Sep 2018


%\cite{Potter:2017ajd}
\bibitem{Potter:2017ajd} 
  C.~T.~Potter,
  %``NMSSM Light Decoupled Higgs Singlet at CEPC,''
  Int.\ J.\ Mod.\ Phys.\ A {\bf 32}, no. 34, 1746011 (2017)
  doi:10.1142/S0217751X17460113
  [arXiv:1707.00041 [hep-ph]].
  %%CITATION = doi:10.1142/S0217751X17460113;%%
  %1 citations counted in INSPIRE as of 09 Sep 2018


%\cite{Fan:2014vta}
\bibitem{Fan:2014vta} 
  J.~Fan, M.~Reece and L.~T.~Wang,
  %``Possible Futures of Electroweak Precision: ILC, FCC-ee, and CEPC,''
  JHEP {\bf 1509}, 196 (2015)
  doi:10.1007/JHEP09(2015)196
  [arXiv:1411.1054 [hep-ph]].
  %%CITATION = doi:10.1007/JHEP09(2015)196;%%
  %49 citations counted in INSPIRE as of 09 Sep 2018


%\cite{Fan:2014axa}
\bibitem{Fan:2014axa} 
  J.~Fan, M.~Reece and L.~T.~Wang,
  %``Precision Natural SUSY at CEPC, FCC-ee, and ILC,''
  JHEP {\bf 1508}, 152 (2015)
  doi:10.1007/JHEP08(2015)152
  [arXiv:1412.3107 [hep-ph]].
  %%CITATION = doi:10.1007/JHEP08(2015)152;%%
  %39 citations counted in INSPIRE as of 09 Sep 2018


%\cite{Fedderke:2015txa}
\bibitem{Fedderke:2015txa} 
  M.~A.~Fedderke, T.~Lin and L.~T.~Wang,
  %``Probing the fermionic Higgs portal at lepton colliders,''
  JHEP {\bf 1604}, 160 (2016)
  doi:10.1007/JHEP04(2016)160
  [arXiv:1506.05465 [hep-ph]].
  %%CITATION = doi:10.1007/JHEP04(2016)160;%%
  %20 citations counted in INSPIRE as of 09 Sep 2018


%\cite{Huang:2015izx}
\bibitem{Huang:2015izx} 
  F.~P.~Huang, P.~H.~Gu, P.~F.~Yin, Z.~H.~Yu and X.~Zhang,
  %``Testing the electroweak phase transition and electroweak baryogenesis at the LHC and a circular electron-positron collider,''
  Phys.\ Rev.\ D {\bf 93}, no. 10, 103515 (2016)
  doi:10.1103/PhysRevD.93.103515
  [arXiv:1511.03969 [hep-ph]].
  %%CITATION = doi:10.1103/PhysRevD.93.103515;%%
  %38 citations counted in INSPIRE as of 09 Sep 2018


%\cite{Ge:2016zro}
\bibitem{Ge:2016zro} 
  S.~F.~Ge, H.~J.~He and R.~Q.~Xiao,
  %``Probing new physics scales from Higgs and electroweak observables at e$^{+}$ e$^{−}$ Higgs factory,''
  JHEP {\bf 1610}, 007 (2016)
  doi:10.1007/JHEP10(2016)007
  [arXiv:1603.03385 [hep-ph]].
  %%CITATION = doi:10.1007/JHEP10(2016)007;%%
  %32 citations counted in INSPIRE as of 09 Sep 2018


%\cite{Durieux:2017rsg}
\bibitem{Durieux:2017rsg} 
  G.~Durieux, C.~Grojean, J.~Gu and K.~Wang,
  %``The leptonic future of the Higgs,''
  JHEP {\bf 1709}, 014 (2017)
  doi:10.1007/JHEP09(2017)014
  [arXiv:1704.02333 [hep-ph]].
  %%CITATION = doi:10.1007/JHEP09(2017)014;%%
  %36 citations counted in INSPIRE as of 09 Sep 2018


%\cite{Gori:2015nqa}
\bibitem{Gori:2015nqa} 
  S.~Gori, J.~Gu and L.~T.~Wang,
  %``The $ Zb\overline{b} $ couplings at future e$^{+}$ e$^{−}$ colliders,''
  JHEP {\bf 1604}, 062 (2016)
  doi:10.1007/JHEP04(2016)062
  [arXiv:1508.07010 [hep-ph]].
  %%CITATION = doi:10.1007/JHEP04(2016)062;%%
  %17 citations counted in INSPIRE as of 09 Sep 2018


%\cite{Craig:2015wwr}
\bibitem{Craig:2015wwr} 
  N.~Craig, J.~Gu, Z.~Liu and K.~Wang,
  %``Beyond Higgs Couplings: Probing the Higgs with Angular Observables at Future e$^{+}$ e$^{−}$ Colliders,''
  JHEP {\bf 1603}, 050 (2016)
  doi:10.1007/JHEP03(2016)050
  [arXiv:1512.06877 [hep-ph]].
  %%CITATION = doi:10.1007/JHEP03(2016)050;%%
  %20 citations counted in INSPIRE as of 09 Sep 2018


%\cite{Li:2016zzh}
\bibitem{Li:2016zzh} 
  G.~Li, Y.~n.~Mao, C.~Zhang and S.~h.~Zhu,
  %``Testing CP violation in the scalar sector at future $e^+e^-$ colliders,''
  Phys.\ Rev.\ D {\bf 95}, no. 3, 035015 (2017)
  doi:10.1103/PhysRevD.95.035015
  [arXiv:1611.08518 [hep-ph]].
  %%CITATION = doi:10.1103/PhysRevD.95.035015;%%
  %4 citations counted in INSPIRE as of 09 Sep 2018


%\cite{DiazCruz:1999xe}
\bibitem{DiazCruz:1999xe} 
  J.~L.~Diaz-Cruz and J.~J.~Toscano,
  %``Lepton flavor violating decays of Higgs bosons beyond the standard model,''
  Phys.\ Rev.\ D {\bf 62}, 116005 (2000)
  doi:10.1103/PhysRevD.62.116005
  [hep-ph/9910233].
  %%CITATION = doi:10.1103/PhysRevD.62.116005;%%
  %131 citations counted in INSPIRE as of 09 Sep 2018


%\cite{ArkaniHamed:2000bq}
\bibitem{ArkaniHamed:2000bq} 
  N.~Arkani-Hamed, L.~J.~Hall, H.~Murayama, D.~Tucker-Smith and N.~Weiner,
  %``Small neutrino masses from supersymmetry breaking,''
  Phys.\ Rev.\ D {\bf 64}, 115011 (2001)
  doi:10.1103/PhysRevD.64.115011
  [hep-ph/0006312].
  %%CITATION = doi:10.1103/PhysRevD.64.115011;%%
  %208 citations counted in INSPIRE as of 09 Sep 2018


%\cite{Perez:2008ee}
\bibitem{Perez:2008ee} 
  G.~Perez and L.~Randall,
  %``Natural Neutrino Masses and Mixings from Warped Geometry,''
  JHEP {\bf 0901}, 077 (2009)
  doi:10.1088/1126-6708/2009/01/077
  [arXiv:0805.4652 [hep-ph]].
  %%CITATION = doi:10.1088/1126-6708/2009/01/077;%%
  %113 citations counted in INSPIRE as of 09 Sep 2018


%\cite{Bjorken:1977vt}
\bibitem{Bjorken:1977vt} 
  J.~D.~Bjorken and S.~Weinberg,
  %``A Mechanism for Nonconservation of Muon Number,''
  Phys.\ Rev.\ Lett.\  {\bf 38}, 622 (1977).
  doi:10.1103/PhysRevLett.38.622
  %%CITATION = doi:10.1103/PhysRevLett.38.622;%%
  %192 citations counted in INSPIRE as of 09 Sep 2018


%\cite{McWilliams:1980kj}
\bibitem{McWilliams:1980kj} 
  B.~McWilliams and L.~F.~Li,
  %``Virtual Effects of Higgs Particles,''
  Nucl.\ Phys.\ B {\bf 179}, 62 (1981).
  doi:10.1016/0550-3213(81)90249-2
  %%CITATION = doi:10.1016/0550-3213(81)90249-2;%%
  %138 citations counted in INSPIRE as of 09 Sep 2018


%\cite{Han:2000jz}
\bibitem{Han:2000jz} 
  T.~Han and D.~Marfatia,
  %``h ---> mu tau at hadron colliders,''
  Phys.\ Rev.\ Lett.\  {\bf 86}, 1442 (2001)
  doi:10.1103/PhysRevLett.86.1442
  [hep-ph/0008141].
  %%CITATION = doi:10.1103/PhysRevLett.86.1442;%%
  %83 citations counted in INSPIRE as of 09 Sep 2018


%\cite{Arganda:2004bz}
\bibitem{Arganda:2004bz} 
  E.~Arganda, A.~M.~Curiel, M.~J.~Herrero and D.~Temes,
  %``Lepton flavor violating Higgs boson decays from massive seesaw neutrinos,''
  Phys.\ Rev.\ D {\bf 71}, 035011 (2005)
  doi:10.1103/PhysRevD.71.035011
  [hep-ph/0407302].
  %%CITATION = doi:10.1103/PhysRevD.71.035011;%%
  %123 citations counted in INSPIRE as of 09 Sep 2018


%\cite{Goudelis:2011un}
\bibitem{Goudelis:2011un} 
  A.~Goudelis, O.~Lebedev and J.~h.~Park,
  %``Higgs-induced lepton flavor violation,''
  Phys.\ Lett.\ B {\bf 707}, 369 (2012)
  doi:10.1016/j.physletb.2011.12.059
  [arXiv:1111.1715 [hep-ph]].
  %%CITATION = doi:10.1016/j.physletb.2011.12.059;%%
  %73 citations counted in INSPIRE as of 09 Sep 2018


%\cite{Arhrib:2012mg}
\bibitem{Arhrib:2012mg} 
  A.~Arhrib, Y.~Cheng and O.~C.~W.~Kong,
  %``Higgs to mu+tau Decay in Supersymmetry without R-parity,''
  EPL {\bf 101}, no. 3, 31003 (2013)
  doi:10.1209/0295-5075/101/31003
  [arXiv:1208.4669 [hep-ph]].
  %%CITATION = doi:10.1209/0295-5075/101/31003;%%
  %37 citations counted in INSPIRE as of 09 Sep 2018


%\cite{Azatov:2009na}
\bibitem{Azatov:2009na} 
  A.~Azatov, M.~Toharia and L.~Zhu,
  %``Higgs Mediated FCNC's in Warped Extra Dimensions,''
  Phys.\ Rev.\ D {\bf 80}, 035016 (2009)
  doi:10.1103/PhysRevD.80.035016
  [arXiv:0906.1990 [hep-ph]].
  %%CITATION = doi:10.1103/PhysRevD.80.035016;%%
  %121 citations counted in INSPIRE as of 09 Sep 2018


%\cite{Blankenburg:2012ex}
\bibitem{Blankenburg:2012ex} 
  G.~Blankenburg, J.~Ellis and G.~Isidori,
  %``Flavour-Changing Decays of a 125 GeV Higgs-like Particle,''
  Phys.\ Lett.\ B {\bf 712}, 386 (2012)
  doi:10.1016/j.physletb.2012.05.007
  [arXiv:1202.5704 [hep-ph]].
  %%CITATION = doi:10.1016/j.physletb.2012.05.007;%%
  %168 citations counted in INSPIRE as of 09 Sep 2018


%\cite{Arganda:2014dta}
\bibitem{Arganda:2014dta} 
  E.~Arganda, M.~J.~Herrero, X.~Marcano and C.~Weiland,
  %``Imprints of massive inverse seesaw model neutrinos in lepton flavor violating Higgs boson decays,''
  Phys.\ Rev.\ D {\bf 91}, no. 1, 015001 (2015)
  doi:10.1103/PhysRevD.91.015001
  [arXiv:1405.4300 [hep-ph]].
  %%CITATION = doi:10.1103/PhysRevD.91.015001;%%
  %71 citations counted in INSPIRE as of 09 Sep 2018


%\cite{Arganda:2015naa}
\bibitem{Arganda:2015naa} 
  E.~Arganda, M.~J.~Herrero, X.~Marcano and C.~Weiland,
  %``Enhancement of the lepton flavor violating Higgs boson decay rates from SUSY loops in the inverse seesaw model,''
  Phys.\ Rev.\ D {\bf 93}, no. 5, 055010 (2016)
  doi:10.1103/PhysRevD.93.055010
  [arXiv:1508.04623 [hep-ph]].
  %%CITATION = doi:10.1103/PhysRevD.93.055010;%%
  %45 citations counted in INSPIRE as of 09 Sep 2018


%\cite{Arganda:2015uca}
\bibitem{Arganda:2015uca} 
  E.~Arganda, M.~J.~Herrero, R.~Morales and A.~Szynkman,
  %``Analysis of the h, H, A → τμ decays induced from SUSY loops within the Mass Insertion Approximation,''
  JHEP {\bf 1603}, 055 (2016)
  doi:10.1007/JHEP03(2016)055
  [arXiv:1510.04685 [hep-ph]].
  %%CITATION = doi:10.1007/JHEP03(2016)055;%%
  %35 citations counted in INSPIRE as of 09 Sep 2018


%\cite{Huang:2015vpt}
\bibitem{Huang:2015vpt} 
  W.~Huang and Y.~L.~Tang,
  %``Flavor anomalies at the LHC and the R-parity violating supersymmetric model extended with vectorlike particles,''
  Phys.\ Rev.\ D {\bf 92}, no. 9, 094015 (2015)
  doi:10.1103/PhysRevD.92.094015
  [arXiv:1509.08599 [hep-ph]].
  %%CITATION = doi:10.1103/PhysRevD.92.094015;%%
  %13 citations counted in INSPIRE as of 09 Sep 2018


%\cite{Crivellin:2015mga}
\bibitem{Crivellin:2015mga} 
  A.~Crivellin, G.~D'Ambrosio and J.~Heeck,
  %``Explaining $h\to\mu^\pm\tau^\mp$, $B\to K^* \mu^+\mu^-$ and $B\to K \mu^+\mu^-/B\to K e^+e^-$ in a two-Higgs-doublet model with gauged $L_\mu-L_\tau$,''
  Phys.\ Rev.\ Lett.\  {\bf 114}, 151801 (2015)
  doi:10.1103/PhysRevLett.114.151801
  [arXiv:1501.00993 [hep-ph]].
  %%CITATION = doi:10.1103/PhysRevLett.114.151801;%%
  %260 citations counted in INSPIRE as of 09 Sep 2018


%\cite{Beneke:2015lba}
\bibitem{Beneke:2015lba} 
  M.~Beneke, P.~Moch and J.~Rohrwild,
  %``Lepton flavour violation in RS models with a brane- or nearly brane-localized Higgs,''
  Nucl.\ Phys.\ B {\bf 906}, 561 (2016)
  doi:10.1016/j.nuclphysb.2016.02.037
  [arXiv:1508.01705 [hep-ph]].
  %%CITATION = doi:10.1016/j.nuclphysb.2016.02.037;%%
  %17 citations counted in INSPIRE as of 09 Sep 2018


%\cite{Baek:2015fma}
\bibitem{Baek:2015fma} 
  S.~Baek and Z.~F.~Kang,
  %``Naturally Large Radiative Lepton Flavor Violating Higgs Decay Mediated by Lepton-flavored Dark Matter,''
  JHEP {\bf 1603}, 106 (2016)
  doi:10.1007/JHEP03(2016)106
  [arXiv:1510.00100 [hep-ph]].
  %%CITATION = doi:10.1007/JHEP03(2016)106;%%
  %43 citations counted in INSPIRE as of 09 Sep 2018


%\cite{Feldmann:2016hvo}
\bibitem{Feldmann:2016hvo} 
  T.~Feldmann, C.~Luhn and P.~Moch,
  %``Lepton-flavour violation in a Pati-Salam model with gauged flavour symmetry,''
  JHEP {\bf 1611}, 078 (2016)
  doi:10.1007/JHEP11(2016)078
  [arXiv:1608.04124 [hep-ph]].
  %%CITATION = doi:10.1007/JHEP11(2016)078;%%
  %9 citations counted in INSPIRE as of 09 Sep 2018


%\cite{Han:2016bvl}
\bibitem{Han:2016bvl} 
  X.~F.~Han, L.~Wang and J.~M.~Yang,
  %``An extension of two-Higgs-doublet model and the excesses of 750 GeV diphoton, muon g-2 and $h\to\mu\tau$,''
  Phys.\ Lett.\ B {\bf 757}, 537 (2016)
  doi:10.1016/j.physletb.2016.04.036
  [arXiv:1601.04954 [hep-ph]].
  %%CITATION = doi:10.1016/j.physletb.2016.04.036;%%
  %63 citations counted in INSPIRE as of 09 Sep 2018


%\cite{Thuc:2016qva}
\bibitem{Thuc:2016qva} 
  T.~T.~Thuc, L.~T.~Hue, H.~N.~Long and T.~P.~Nguyen,
  %``Lepton flavor violating decay of SM-like Higgs boson in a radiative neutrino mass model,''
  Phys.\ Rev.\ D {\bf 93}, no. 11, 115026 (2016)
  doi:10.1103/PhysRevD.93.115026
  [arXiv:1604.03285 [hep-ph]].
  %%CITATION = doi:10.1103/PhysRevD.93.115026;%%
  %15 citations counted in INSPIRE as of 09 Sep 2018


%\cite{Khachatryan:2015kon}
\bibitem{Khachatryan:2015kon} 
  V.~Khachatryan {\it et al.} [CMS Collaboration],
  %``Search for Lepton-Flavour-Violating Decays of the Higgs Boson,''
  Phys.\ Lett.\ B {\bf 749}, 337 (2015)
  doi:10.1016/j.physletb.2015.07.053
  [arXiv:1502.07400 [hep-ex]].
  %%CITATION = doi:10.1016/j.physletb.2015.07.053;%%
  %270 citations counted in INSPIRE as of 09 Sep 2018


%\cite{Aad:2016blu}
\bibitem{Aad:2016blu} 
  G.~Aad {\it et al.} [ATLAS Collaboration],
  %``Search for lepton-flavour-violating decays of the Higgs and $Z$ bosons with the ATLAS detector,''
  Eur.\ Phys.\ J.\ C {\bf 77}, no. 2, 70 (2017)
  doi:10.1140/epjc/s10052-017-4624-0
  [arXiv:1604.07730 [hep-ex]].
  %%CITATION = doi:10.1140/epjc/s10052-017-4624-0;%%
  %84 citations counted in INSPIRE as of 09 Sep 2018


%\cite{CMS:2016qvi}
\bibitem{CMS:2016qvi} 
  CMS Collaboration [CMS Collaboration],
  %``Search for Lepton Flavour Violating Decays of the Higgs Boson in the mu-tau final state at 13 TeV,''
  CMS-PAS-HIG-16-005.
  %%CITATION = CMS-PAS-HIG-16-005;%%
  %42 citations counted in INSPIRE as of 09 Sep 2018


%\cite{Lees:2012xj}
\bibitem{Lees:2012xj} 
  J.~P.~Lees {\it et al.} [BaBar Collaboration],
  %``Evidence for an excess of $\bar{B} \to D^{(*)} \tau^-\bar{\nu}_\tau$ decays,''
  Phys.\ Rev.\ Lett.\  {\bf 109}, 101802 (2012)
  doi:10.1103/PhysRevLett.109.101802
  [arXiv:1205.5442 [hep-ex]].
  %%CITATION = doi:10.1103/PhysRevLett.109.101802;%%
  %580 citations counted in INSPIRE as of 09 Sep 2018


%\cite{Aaij:2014ora}
\bibitem{Aaij:2014ora} 
  R.~Aaij {\it et al.} [LHCb Collaboration],
  %``Test of lepton universality using $B^{+}\rightarrow K^{+}\ell^{+}\ell^{-}$ decays,''
  Phys.\ Rev.\ Lett.\  {\bf 113}, 151601 (2014)
  doi:10.1103/PhysRevLett.113.151601
  [arXiv:1406.6482 [hep-ex]].
  %%CITATION = doi:10.1103/PhysRevLett.113.151601;%%
  %657 citations counted in INSPIRE as of 09 Sep 2018


%\cite{Huschle:2015rga}
\bibitem{Huschle:2015rga} 
  M.~Huschle {\it et al.} [Belle Collaboration],
  %``Measurement of the branching ratio of $\bar{B} \to D^{(\ast)} \tau^- \bar{\nu}_\tau$ relative to $\bar{B} \to D^{(\ast)} \ell^- \bar{\nu}_\ell$ decays with hadronic tagging at Belle,''
  Phys.\ Rev.\ D {\bf 92}, no. 7, 072014 (2015)
  doi:10.1103/PhysRevD.92.072014
  [arXiv:1507.03233 [hep-ex]].
  %%CITATION = doi:10.1103/PhysRevD.92.072014;%%
  %385 citations counted in INSPIRE as of 09 Sep 2018


%\cite{Aaij:2017vbb}
\bibitem{Aaij:2017vbb} 
  R.~Aaij {\it et al.} [LHCb Collaboration],
  %``Test of lepton universality with $B^{0} \rightarrow K^{*0}\ell^{+}\ell^{-}$ decays,''
  JHEP {\bf 1708}, 055 (2017)
  doi:10.1007/JHEP08(2017)055
  [arXiv:1705.05802 [hep-ex]].
  %%CITATION = doi:10.1007/JHEP08(2017)055;%%
  %259 citations counted in INSPIRE as of 09 Sep 2018


%\cite{Banerjee:2016foh}
\bibitem{Banerjee:2016foh} 
  S.~Banerjee, B.~Bhattacherjee, M.~Mitra and M.~Spannowsky,
  %``The Lepton Flavour Violating Higgs Decays at the HL-LHC and the ILC,''
  JHEP {\bf 1607}, 059 (2016)
  doi:10.1007/JHEP07(2016)059
  [arXiv:1603.05952 [hep-ph]].
  %%CITATION = doi:10.1007/JHEP07(2016)059;%%
  %24 citations counted in INSPIRE as of 09 Sep 2018


%\cite{Kanemura:2004cn}
\bibitem{Kanemura:2004cn} 
  S.~Kanemura, K.~Matsuda, T.~Ota, T.~Shindou, E.~Takasugi and K.~Tsumura,
  %``Search for lepton flavor violation in the Higgs boson decay at a linear collider,''
  Phys.\ Lett.\ B {\bf 599}, 83 (2004)
  doi:10.1016/j.physletb.2004.08.024
  [hep-ph/0406316].
  %%CITATION = doi:10.1016/j.physletb.2004.08.024;%%
  %55 citations counted in INSPIRE as of 09 Sep 2018


%\cite{Chakraborty:2016gff}
\bibitem{Chakraborty:2016gff} 
  I.~Chakraborty, A.~Datta and A.~Kundu,
  %``Lepton flavor violating Higgs boson decay ${\boldsymbol{h}} \rightarrow \mu \tau $ at the ILC,''
  J.\ Phys.\ G {\bf 43}, no. 12, 125001 (2016)
  doi:10.1088/0954-3899/43/12/125001
  [arXiv:1603.06681 [hep-ph]].
  %%CITATION = doi:10.1088/0954-3899/43/12/125001;%%
  %14 citations counted in INSPIRE as of 09 Sep 2018


%\cite{Chakraborty:2017tyb}
\bibitem{Chakraborty:2017tyb} 
  I.~Chakraborty, S.~Mondal and B.~Mukhopadhyaya,
  %``Lepton flavor violating Higgs boson decay at $e^+ e^-$ colliders,''
  Phys.\ Rev.\ D {\bf 96}, no. 11, 115020 (2017)
  doi:10.1103/PhysRevD.96.115020
  [arXiv:1709.08112 [hep-ph]].
  %%CITATION = doi:10.1103/PhysRevD.96.115020;%%
  %3 citations counted in INSPIRE as of 09 Sep 2018


%\cite{Alwall:2014hca}
\bibitem{Alwall:2014hca} 
  J.~Alwall {\it et al.},
  %``The automated computation of tree-level and next-to-leading order differential cross sections, and their matching to parton shower simulations,''
  JHEP {\bf 1407}, 079 (2014)
  doi:10.1007/JHEP07(2014)079
  [arXiv:1405.0301 [hep-ph]].
  %%CITATION = doi:10.1007/JHEP07(2014)079;%%
  %3053 citations counted in INSPIRE as of 09 Sep 2018


%\cite{Moretti:2001zz}
\bibitem{Moretti:2001zz} 
  M.~Moretti, T.~Ohl and J.~Reuter,
  %``O'Mega: An Optimizing matrix element generator,''
  hep-ph/0102195.
  %%CITATION = HEP-PH/0102195;%%
  %346 citations counted in INSPIRE as of 09 Sep 2018


%\cite{Kilian:2007gr}
\bibitem{Kilian:2007gr} 
  W.~Kilian, T.~Ohl and J.~Reuter,
  %``WHIZARD: Simulating Multi-Particle Processes at LHC and ILC,''
  Eur.\ Phys.\ J.\ C {\bf 71}, 1742 (2011)
  doi:10.1140/epjc/s10052-011-1742-y
  [arXiv:0708.4233 [hep-ph]].
  %%CITATION = doi:10.1140/epjc/s10052-011-1742-y;%%
  %518 citations counted in INSPIRE as of 09 Sep 2018


%\cite{Sjostrand:2006za}
\bibitem{Sjostrand:2006za} 
  T.~Sjostrand, S.~Mrenna and P.~Z.~Skands,
  %``PYTHIA 6.4 Physics and Manual,''
  JHEP {\bf 0605}, 026 (2006)
  doi:10.1088/1126-6708/2006/05/026
  [hep-ph/0603175].
  %%CITATION = doi:10.1088/1126-6708/2006/05/026;%%
  %9662 citations counted in INSPIRE as of 09 Sep 2018


%\cite{Cacciari:2005hq}
\bibitem{Cacciari:2005hq} 
  M.~Cacciari and G.~P.~Salam,
  %``Dispelling the $N^{3}$ myth for the $k_t$ jet-finder,''
  Phys.\ Lett.\ B {\bf 641}, 57 (2006)
  doi:10.1016/j.physletb.2006.08.037
  [hep-ph/0512210].
  %%CITATION = doi:10.1016/j.physletb.2006.08.037;%%
  %1681 citations counted in INSPIRE as of 09 Sep 2018


%\cite{Cacciari:2011ma}
\bibitem{Cacciari:2011ma} 
  M.~Cacciari, G.~P.~Salam and G.~Soyez,
  %``FastJet User Manual,''
  Eur.\ Phys.\ J.\ C {\bf 72}, 1896 (2012)
  doi:10.1140/epjc/s10052-012-1896-2
  [arXiv:1111.6097 [hep-ph]].
  %%CITATION = doi:10.1140/epjc/s10052-012-1896-2;%%
  %2500 citations counted in INSPIRE as of 09 Sep 2018


%\cite{Bardin:1994sc}
\bibitem{Bardin:1994sc} 
  D.~Y.~Bardin, M.~S.~Bilenky, D.~Lehner, A.~Olchevski and T.~Riemann,
  %``Semi-analytical approach to four-fermion production in $e^+ e^−$ annihilation,''
  Nucl.\ Phys.\ Proc.\ Suppl.\  {\bf 37B}, no. 2, 148 (1994)
  doi:10.1016/0920-5632(94)90670-X
  [hep-ph/9406340].
  %%CITATION = doi:10.1016/0920-5632(94)90670-X;%%
  %68 citations counted in INSPIRE as of 09 Sep 2018


%\cite{Olive:2016xmw}
\bibitem{Olive:2016xmw} 
  C.~Patrignani {\it et al.} [Particle Data Group],
  %``Review of Particle Physics,''
  Chin.\ Phys.\ C {\bf 40}, no. 10, 100001 (2016).
  doi:10.1088/1674-1137/40/10/100001
  %%CITATION = doi:10.1088/1674-1137/40/10/100001;%%
  %4090 citations counted in INSPIRE as of 09 Sep 2018


%\cite{Khachatryan:2016rke}
\bibitem{Khachatryan:2016rke} 
  V.~Khachatryan {\it et al.} [CMS Collaboration],
  %``Search for lepton flavour violating decays of the Higgs boson to $e \tau$ and $e \mu$ in proton–proton collisions at $\sqrt s=$ 8 TeV,''
  Phys.\ Lett.\ B {\bf 763}, 472 (2016)
  doi:10.1016/j.physletb.2016.09.062
  [arXiv:1607.03561 [hep-ex]].
  %%CITATION = doi:10.1016/j.physletb.2016.09.062;%%
  %36 citations counted in INSPIRE as of 09 Sep 2018


%\cite{Kawada:2015wea}
\bibitem{Kawada:2015wea} 
  S.~i.~Kawada, K.~Fujii, T.~Suehara, T.~Takahashi and T.~Tanabe,
  %``A study of the measurement precision of the Higgs boson decaying into tau pairs at the ILC,''
  Eur.\ Phys.\ J.\ C {\bf 75}, no. 12, 617 (2015)
  doi:10.1140/epjc/s10052-015-3854-2
  [arXiv:1509.01885 [hep-ex]].
  %%CITATION = doi:10.1140/epjc/s10052-015-3854-2;%%
  %4 citations counted in INSPIRE as of 09 Sep 2018


%\cite{Denner:2011mq}
\bibitem{Denner:2011mq} 
  A.~Denner, S.~Heinemeyer, I.~Puljak, D.~Rebuzzi and M.~Spira,
  %``Standard Model Higgs-Boson Branching Ratios with Uncertainties,''
  Eur.\ Phys.\ J.\ C {\bf 71}, 1753 (2011)
  doi:10.1140/epjc/s10052-011-1753-8
  [arXiv:1107.5909 [hep-ph]].
  %%CITATION = doi:10.1140/epjc/s10052-011-1753-8;%%
  %230 citations counted in INSPIRE as of 09 Sep 2018


%\cite{Buchmuller:1985jz}
\bibitem{Buchmuller:1985jz} 
  W.~Buchmuller and D.~Wyler,
  %``Effective Lagrangian Analysis of New Interactions and Flavor Conservation,''
  Nucl.\ Phys.\ B {\bf 268}, 621 (1986).
  doi:10.1016/0550-3213(86)90262-2
  %%CITATION = doi:10.1016/0550-3213(86)90262-2;%%
  %1428 citations counted in INSPIRE as of 09 Sep 2018


%\cite{Grzadkowski:2010es}
\bibitem{Grzadkowski:2010es} 
  B.~Grzadkowski, M.~Iskrzynski, M.~Misiak and J.~Rosiek,
  %``Dimension-Six Terms in the Standard Model Lagrangian,''
  JHEP {\bf 1010}, 085 (2010)
  doi:10.1007/JHEP10(2010)085
  [arXiv:1008.4884 [hep-ph]].
  %%CITATION = doi:10.1007/JHEP10(2010)085;%%
  %746 citations counted in INSPIRE as of 09 Sep 2018


%\cite{Harnik:2012pb}
\bibitem{Harnik:2012pb} 
  R.~Harnik, J.~Kopp and J.~Zupan,
  %``Flavor Violating Higgs Decays,''
  JHEP {\bf 1303}, 026 (2013)
  doi:10.1007/JHEP03(2013)026
  [arXiv:1209.1397 [hep-ph]].
  %%CITATION = doi:10.1007/JHEP03(2013)026;%%
  %246 citations counted in INSPIRE as of 09 Sep 2018


%\cite{Alonso:2013hga}
\bibitem{Alonso:2013hga} 
  R.~Alonso, E.~E.~Jenkins, A.~V.~Manohar and M.~Trott,
  %``Renormalization Group Evolution of the Standard Model Dimension Six Operators III: Gauge Coupling Dependence and Phenomenology,''
  JHEP {\bf 1404}, 159 (2014)
  doi:10.1007/JHEP04(2014)159
  [arXiv:1312.2014 [hep-ph]].
  %%CITATION = doi:10.1007/JHEP04(2014)159;%%
  %249 citations counted in INSPIRE as of 09 Sep 2018


%\cite{Pruna:2014asa}
\bibitem{Pruna:2014asa} 
  G.~M.~Pruna and A.~Signer,
  %``The $\mu\to e\gamma$ decay in a systematic effective field theory approach with dimension 6 operators,''
  JHEP {\bf 1410}, 014 (2014)
  doi:10.1007/JHEP10(2014)014
  [arXiv:1408.3565 [hep-ph]].
  %%CITATION = doi:10.1007/JHEP10(2014)014;%%
  %36 citations counted in INSPIRE as of 09 Sep 2018


%\cite{Pruna:2015jhf}
\bibitem{Pruna:2015jhf} 
  G.~M.~Pruna and A.~Signer,
  %``Lepton-flavour violating decays in theories with dimension 6 operators,''
  EPJ Web Conf.\  {\bf 118}, 01031 (2016)
  doi:10.1051/epjconf/201611801031
  [arXiv:1511.04421 [hep-ph]].
  %%CITATION = doi:10.1051/epjconf/201611801031;%%
  %12 citations counted in INSPIRE as of 09 Sep 2018


%\cite{Cheng:1987rs}
\bibitem{Cheng:1987rs} 
  T.~P.~Cheng and M.~Sher,
  %``Mass Matrix Ansatz and Flavor Nonconservation in Models with Multiple Higgs Doublets,''
  Phys.\ Rev.\ D {\bf 35}, 3484 (1987).
  doi:10.1103/PhysRevD.35.3484
  %%CITATION = doi:10.1103/PhysRevD.35.3484;%%
  %539 citations counted in INSPIRE as of 09 Sep 2018


%\cite{Lee:1973iz}
\bibitem{Lee:1973iz} 
  T.~D.~Lee,
  %``A Theory of Spontaneous T Violation,''
  Phys.\ Rev.\ D {\bf 8}, 1226 (1973).
  doi:10.1103/PhysRevD.8.1226
  %%CITATION = doi:10.1103/PhysRevD.8.1226;%%
  %1169 citations counted in INSPIRE as of 09 Sep 2018


%\cite{Haber:1984rc}
\bibitem{Haber:1984rc} 
  H.~E.~Haber and G.~L.~Kane,
  %``The Search for Supersymmetry: Probing Physics Beyond the Standard Model,''
  Phys.\ Rept.\  {\bf 117}, 75 (1985).
  doi:10.1016/0370-1573(85)90051-1
  %%CITATION = doi:10.1016/0370-1573(85)90051-1;%%
  %4887 citations counted in INSPIRE as of 09 Sep 2018


%\cite{Kim:1986ax}
\bibitem{Kim:1986ax} 
  J.~E.~Kim,
  %``Light Pseudoscalars, Particle Physics and Cosmology,''
  Phys.\ Rept.\  {\bf 150}, 1 (1987).
  doi:10.1016/0370-1573(87)90017-2
  %%CITATION = doi:10.1016/0370-1573(87)90017-2;%%
  %1055 citations counted in INSPIRE as of 09 Sep 2018


%\cite{Turok:1990zg}
\bibitem{Turok:1990zg} 
  N.~Turok and J.~Zadrozny,
  %``Electroweak baryogenesis in the two doublet model,''
  Nucl.\ Phys.\ B {\bf 358}, 471 (1991).
  doi:10.1016/0550-3213(91)90356-3
  %%CITATION = doi:10.1016/0550-3213(91)90356-3;%%
  %239 citations counted in INSPIRE as of 09 Sep 2018


%\cite{Branco:2011iw}
\bibitem{Branco:2011iw} 
  G.~C.~Branco, P.~M.~Ferreira, L.~Lavoura, M.~N.~Rebelo, M.~Sher and J.~P.~Silva,
  %``Theory and phenomenology of two-Higgs-doublet models,''
  Phys.\ Rept.\  {\bf 516}, 1 (2012)
  doi:10.1016/j.physrep.2012.02.002
  [arXiv:1106.0034 [hep-ph]].
  %%CITATION = doi:10.1016/j.physrep.2012.02.002;%%
  %1295 citations counted in INSPIRE as of 09 Sep 2018


%\cite{Aaboud:2018zhk}
\bibitem{Aaboud:2018zhk} 
  M.~Aaboud {\it et al.} [ATLAS Collaboration],
  %``Observation of $H \rightarrow b\bar{b}$ decays and $VH$ production with the ATLAS detector,''
  arXiv:1808.08238 [hep-ex].
  %%CITATION = ARXIV:1808.08238;%%


%\cite{Mariotti:2016owy}
\bibitem{Mariotti:2016owy} 
  C.~Mariotti and G.~Passarino,
  %``Higgs boson couplings: measurements and theoretical interpretation,''
  Int.\ J.\ Mod.\ Phys.\ A {\bf 32}, no. 04, 1730003 (2017)
  doi:10.1142/S0217751X17300034
  [arXiv:1612.00269 [hep-ph]].
  %%CITATION = doi:10.1142/S0217751X17300034;%%
  %12 citations counted in INSPIRE as of 09 Sep 2018


%\cite{Sirunyan:2018kst}
\bibitem{Sirunyan:2018kst} 
  A.~M.~Sirunyan {\it et al.} [CMS Collaboration],
  %``Observation of Higgs boson decay to bottom quarks,''
  %Submitted to: Phys.Rev.Lett.
  [arXiv:1808.08242 [hep-ex]].
  %%CITATION = ARXIV:1808.08242;%%


%\cite{Randall:1999ee}
\bibitem{Randall:1999ee} 
  L.~Randall and R.~Sundrum,
  %``A Large mass hierarchy from a small extra dimension,''
  Phys.\ Rev.\ Lett.\  {\bf 83}, 3370 (1999)
  doi:10.1103/PhysRevLett.83.3370
  [hep-ph/9905221].
  %%CITATION = doi:10.1103/PhysRevLett.83.3370;%%
  %8068 citations counted in INSPIRE as of 09 Sep 2018


%\cite{Randall:1999vf}
\bibitem{Randall:1999vf} 
  L.~Randall and R.~Sundrum,
  %``An Alternative to compactification,''
  Phys.\ Rev.\ Lett.\  {\bf 83}, 4690 (1999)
  doi:10.1103/PhysRevLett.83.4690
  [hep-th/9906064].
  %%CITATION = doi:10.1103/PhysRevLett.83.4690;%%
  %6261 citations counted in INSPIRE as of 09 Sep 2018


%\cite{Gherghetta:2000qt}
\bibitem{Gherghetta:2000qt} 
  T.~Gherghetta and A.~Pomarol,
  %``Bulk fields and supersymmetry in a slice of AdS,''
  Nucl.\ Phys.\ B {\bf 586}, 141 (2000)
  doi:10.1016/S0550-3213(00)00392-8
  [hep-ph/0003129].
  %%CITATION = doi:10.1016/S0550-3213(00)00392-8;%%
  %1112 citations counted in INSPIRE as of 09 Sep 2018


%\cite{Grossman:1999ra}
\bibitem{Grossman:1999ra} 
  Y.~Grossman and M.~Neubert,
  %``Neutrino masses and mixings in nonfactorizable geometry,''
  Phys.\ Lett.\ B {\bf 474}, 361 (2000)
  doi:10.1016/S0370-2693(00)00054-X
  [hep-ph/9912408].
  %%CITATION = doi:10.1016/S0370-2693(00)00054-X;%%
  %852 citations counted in INSPIRE as of 09 Sep 2018


%\cite{Frank:2014aca}
\bibitem{Frank:2014aca} 
  M.~Frank, C.~Hamzaoui, N.~Pourtolami and M.~Toharia,
  %``Unified Flavor Symmetry from warped dimensions,''
  Phys.\ Lett.\ B {\bf 742}, 178 (2015)
  doi:10.1016/j.physletb.2015.01.025
  [arXiv:1406.2331 [hep-ph]].
  %%CITATION = doi:10.1016/j.physletb.2015.01.025;%%
  %13 citations counted in INSPIRE as of 09 Sep 2018


%\cite{Casagrande:2008hr}
\bibitem{Casagrande:2008hr} 
  S.~Casagrande, F.~Goertz, U.~Haisch, M.~Neubert and T.~Pfoh,
  %``Flavor Physics in the Randall-Sundrum Model: I. Theoretical Setup and Electroweak Precision Tests,''
  JHEP {\bf 0810}, 094 (2008)
  doi:10.1088/1126-6708/2008/10/094
  [arXiv:0807.4937 [hep-ph]].
  %%CITATION = doi:10.1088/1126-6708/2008/10/094;%%
  %240 citations counted in INSPIRE as of 09 Sep 2018


%\cite{Kaluza:1921tu}
\bibitem{Kaluza:1921tu} 
  T.~Kaluza,
  %``Zum Unitätsproblem der Physik,''
  Sitzungsber.\ Preuss.\ Akad.\ Wiss.\ Berlin (Math.\ Phys.\ ) {\bf 1921}, 966 (1921)
  [arXiv:1803.08616 [physics.hist-ph]].
  %%CITATION = ARXIV:1803.08616;%%
  %672 citations counted in INSPIRE as of 09 Sep 2018


%\cite{Klein:1926tv}
\bibitem{Klein:1926tv} 
  O.~Klein,
  %``Quantum Theory and Five-Dimensional Theory of Relativity. (In German and English),''
  Z.\ Phys.\  {\bf 37}, 895 (1926)
  [Surveys High Energ.\ Phys.\  {\bf 5}, 241 (1986)].
  doi:10.1007/BF01397481
  %%CITATION = doi:10.1007/BF01397481;%%
  %2173 citations counted in INSPIRE as of 09 Sep 2018


%\cite{Ilakovac:1994kj}
\bibitem{Ilakovac:1994kj} 
  A.~Ilakovac and A.~Pilaftsis,
  %``Flavor violating charged lepton decays in seesaw-type models,''
  Nucl.\ Phys.\ B {\bf 437}, 491 (1995)
  doi:10.1016/0550-3213(94)00567-X
  [hep-ph/9403398].
  %%CITATION = doi:10.1016/0550-3213(94)00567-X;%%
  %298 citations counted in INSPIRE as of 09 Sep 2018


%\cite{Arganda:2017vdb}
\bibitem{Arganda:2017vdb} 
  E.~Arganda, M.~J.~Herrero, X.~Marcano, R.~Morales and A.~Szynkman,
  %``Effective lepton flavor violating Hℓiℓj vertex from right-handed neutrinos within the mass insertion approximation,''
  Phys.\ Rev.\ D {\bf 95}, no. 9, 095029 (2017)
  doi:10.1103/PhysRevD.95.095029
  [arXiv:1612.09290 [hep-ph]].
  %%CITATION = doi:10.1103/PhysRevD.95.095029;%%
  %9 citations counted in INSPIRE as of 09 Sep 2018


%\cite{Marcano:2017ucg}
\bibitem{Marcano:2017ucg} 
  X.~Marcano Imaz,
  %``Lepton flavor violation from low scale seesaw neutrinos with masses reachable at the LHC,''
  doi:10.1007/978-3-319-94604-7
  arXiv:1710.08032 [hep-ph].
  %%CITATION = doi:10.1007/978-3-319-94604-7;%%
  %1 citations counted in INSPIRE as of 09 Sep 2018
  
    
  \end{thebibliography}
\end{document}